\documentclass[12pt,a4paper]{article}
\usepackage[utf8]{inputenc}
\usepackage[english]{babel}

\usepackage[margin=1in]{geometry}

\usepackage[sort&compress]{natbib}
\bibpunct{(}{)}{;}{a}{}{,} 

\usepackage[dvipsnames]{xcolor}

\usepackage{amsmath, hyperref, amsfonts, amssymb, amsthm, mathtools, mathrsfs, cleveref}

\usepackage{microtype}
\usepackage{fullpage}
\usepackage{titlesec}

\titleformat{\section}
  {\normalfont\large\bfseries}{\thesection}{0.8em}{}     
\titleformat{\subsection}
  {\normalfont\bfseries}{\thesubsection}{0.8em}{}       

\usepackage{graphicx}
\usepackage{tikz}
\usepackage{float}
\usetikzlibrary{arrows}
\usepackage{pgfplots}
\pgfplotsset{compat=1.18}
\usepackage{subfigure}
\usepackage{hyperref}

\usepackage{booktabs, multirow, multicol}

\usepackage{algorithm}
\usepackage{algpseudocode}

\theoremstyle{plain} 
\newtheorem{theorem}{Theorem}[section]

\theoremstyle{definition}

\theoremstyle{remark}

\newcommand{\PP}{\mathbb{P}}
\newcommand{\RR}{\mathbb{R}}

\newcommand{\nm}{\mathsf{N}}

\title{Rényi's $\alpha$-divergence variational Bayes for spike-and-slab high-dimensional linear regression}

\author{
    Chadi Bsila\thanks{Department of Statistical Sciences, Wake Forest University, \texttt{bsilc26@wfu.edu}}
    \and
    Yiqi Tang\thanks{Department of Statistics, Colby College, \texttt{ytang@colby.edu}}
    \and
    Kaiwen Wang\thanks{Department of Data Science, Davidson College, \texttt{kewang2@davidson.edu}}
    \thanks{The authors are grateful for the support of Laurie J. Heyer and the Davidson Research Initiative (DRI).}
} 
\date{\today}
\begin{document}
\maketitle
\begin{abstract}
Sparse high-dimensional linear regression is a central problem in statistics, where the goal is often variable selection and/or coefficient estimation. We propose a mean-field variational Bayes approximation for sparse regression with spike-and-slab Laplace priors that replaces the standard Kullback–Leibler (KL) divergence objective with the Rényi's $\alpha$ divergence: a one-parameter generalization of KL divergence indexed by $\alpha \in (0, \infty) \setminus \{1\}$ that allows flexibility between zero-forcing and mass-covering behavior. We derive coordinate ascent variational inference (CAVI) updates via a second-order delta method and develop a stochastic variational inference algorithm based on a Monte Carlo surrogate Rényi lower bound. In simulations, our two methods perform comparably to state-of-the-art Bayesian variable selection procedures across a range of sparsity configurations and $\alpha$ values for both variable selection and estimation, and our numerical results illustrate how different choices of $\alpha$ can be advantageous in different sparsity configurations.  
\end{abstract}
\vspace{1mm}
\section{Introduction}
\label{S:intro}

We consider the familiar linear regression model 
\begin{equation}
    y = X\theta + z, \label{eq:data_model}
\end{equation}
where $y \in \RR^n$ is the $n\times 1$ observed response vector, $X$ is a given, deterministic $n \times p$ design matrix, $\theta \in \RR^p$ is the vector of regression coefficients, and $z \sim \nm_n(0, I_n)$ denotes independent Gaussian noise. We are particularly interested in the high-dimensional setting, where $p\gg n$. Specifically, we focus on variable selection and parameter estimation. 

This common but challenging problem cannot be solved without additional structure; we assume that the coefficient vector $\theta$ is sparse, where most of the entries in $\theta$ are zero. This assumption is common in the literature, and many, both frequentists and Bayesians, have investigated the sparse high-dimensional linear regression problem. In the frequentist realm, the most widely-used method is lasso \citep{tibshirani1996regression} and variants of lasso such as the adaptive lasso \citep{zou2006adaptive}.

On the Bayesian front, the sparsity structure is typically incorporated into the prior distribution to solve high-dimensional problems. The two popular types of priors are spike-and-slab mixture priors \citep{george1993variable, castillo2015bayesian, castillo2012needles} and continuous shrinkage priors like the horseshoe \citep{piironen2017sparsity, carvalho2010horseshoe}. One main benefit of using a Bayesian framework in such problems is a readily available posterior probability distribution for uncertainty quantification; however, a drawback of Bayesian methods, especially in the high-dimensional realm or, even more so, the ultra-high-dimensional settings, is the time and resources the computations require. Bayesian solutions to problems are often more computationally-intensive than their frequentist counterparts to the same problems. This is mainly because most Bayesian methods require Markov Chain Monte Carlo (MCMC) to get to an approximate posterior distribution. This becomes challenging in high-dimensional problems, especially as the number of variables $p$ increases to be in the thousands. A typical Bayesian method that uses MCMC may take up to minutes to compute when $p$ is in the hundreds, and may be too slow to use for $p$ in the thousands. 

For computational efficiency, many now use alternatives to MCMC instead; variational inference (VI) \citep{blei2017variational} is one such popular alternative. Rather than doing integration like MCMC, VI converts the problem into optimization, making it a more efficient way to find a solution. In VI, we first posit a family of models and then find a member of the family that minimizes the Kullback-Leibler (KL) divergence between that member and the original posterior of interest (that is difficult to compute). The method is also referred to as variational approximation, since we are seeking out the family member that would best approximate the posterior distribution. In traditional VI, the best approximation is measured by the smallest KL-divergence between the family member and the posterior distribution; of course, there are alternative measures of distance, and those can be used instead for variational Bayes.  \citet{minka2005divergence} outlines a number of different divergences including Rényi's $\alpha$-divergences and $f$-divergences, and how they may be applied to message passing, a version of variational approximation. 

In our work, we focus on Rényi's $\alpha$-divergence \citep{renyi1961measures}, a flexible measure of distance that exhibits different behavior as its hyperparameter $\alpha$ changes. It can be considered as an extension of the traditional VI with KL-divergence: as $\alpha\rightarrow 1$, Rényi's $\alpha$-divergence is equivalent to KL-divergence. VI using Rényi's $\alpha$-divergence instead of the traditional KL-divergence has been studied in a few different contexts, but has not been widely studied like the KL-divergence VI has been. \citet{li2016renyi} used Rényi's $\alpha$-divergence by proposing a variational Rényi bound, comparable to the Evidence Lower-bound (ELBO), a bound used in the KL-divergence VI, and applied their method to Bayesian neural networks and variational auto-encoders in numerical experiments. Others have also mostly focused on complex models such as deep generative models and Bayesian neural networks \citep[e.g.,][]{bui2016black, liu2021alpha}. 

Different $\alpha$ values exhibit different behaviors, and a better understanding of those behaviors is needed for different problems \citep{li2016renyi}. Our aim is to provide a solution to this question for a common and important setting, the sparse high-dimensional linear regression. We adopt the Rényi's $\alpha$-divergence as the variational objective and investigate how varying $\alpha$ influences variable selection and parameter estimation. Specifically, we use a Laplace spike-and-slab model and use a mean-field family for the variational family. Compared to using KL-divergence, having the $\alpha$ hyperparameter provides more flexibility, but at the same time, the additional flexibility also makes our objective harder to optimize and more difficult to bound. To overcome this obstacle, we used techniques including the multivariate Delta method and Jensen's inequality. We were able to successfully derive a coordinate-ascent algorithm (CAVI), commonly used for VI. As a comparison, we also derive a second algorithm based on the variational Rényi (VR) bound proposed in \citet{li2016renyi}, and empirically investigate how varying $\alpha$ affects parameter estimation and variable selection for our two methods. 

The remainder of the paper is organized as follows. Section~\ref{S:background} reviews VI, with emphasis on the Rényi's $\alpha$-divergence formulation, and describes the hierarchical prior for the Laplace spike-and-slab model, highlighting its sparsity-inducing properties.  Section \ref{S:cavi} derives CAVI updates. Section \ref{S:mc_vb} develops stochastic VI updates based on a Monte Carlo variational Rényi bound: an analogue of the evidence lower bound (ELBO) used in standard VI. Section \ref{S:simulations} presents numerical experiments showing the empirical performance of our two methods as $\alpha$ varies, and compares our two methods to existing Bayesian variable selection procedures. Finally, Section \ref{S:discussion} discusses implications and directions for future work.

\section{Background}
\label{S:background}

\subsection{Notation and preliminaries}

We first define some notations that will be used throughout the paper. For a probability measure $P$ absolutely continuous with respect to a measure $Q$, we write $\frac{dP}{dQ}$ for the Radon–Nikodym derivative. For a vector $v \in \RR^d$, we use $\| v \|_2$ to denote its Euclidean ($\ell_2$) norm. We use the shorthand $D_\alpha$ for $D_\alpha(Q \,\|\, P)$ to denote Rényi's $\alpha$-divergence.

\subsection{Variational inference for high-dimensional linear regression}
As is typical, we employ a hierarchical structure for the prior. Write $\theta = (\theta_S, S)$, where $S \subseteq [p]$ denotes the set of indices corresponding to nonzero coefficients (active signals), and $\theta_S$ the subvector of $\theta$ indexed by $S$. Equivalently, we may represent $S$ by a binary inclusion vector $z \in \{0,1\}^p$ with $z_i = 1$ if $i \in S$ and $z_i = 0$ otherwise. From a Bayesian perspective, sparsity is induced through a model selection prior that assigns positive prior mass to subsets $S \subseteq [p]$. Without VI, posterior approximation requires a combinatorial search over all (or most of) $2^p$ possible models, which becomes computationally infeasible for large $p$. 

Following \cite{ray2022variational}, let $s \in \mathbb{N}$ denote the model size or sparsity level, and consider the hierarchical Laplace spike-and-slab prior:
\begin{align*}
s &\sim \pi(s), \\
S \,\big|\, |S| = s &\sim \mathrm{Unif}_{p,s}, \\
\theta_i \,\big|\, S &\stackrel{\mathrm{ind}}{\sim} 
\begin{cases}
\mathrm{Lap}(\lambda), & \lambda > 0, \quad i \in S, \\
\delta_0, & i \notin S,
\end{cases}
\end{align*}
where $\mathrm{Lap}(\lambda)$ denotes the Laplace distribution with rate $\lambda$ and $\delta_0$ the Dirac measure at zero.  

The prior $\pi(s)$ is chosen to induce sparsity by underweighting large models while placing sufficient mass near the true model size. We assume constants $A_1, A_2, A_3, A_4 > 0$ exist such that
\begin{align}
A_1 p^{-A_3} \,\pi_p(s-1) \ \leq\  \pi_p(s) \ \leq\ A_2 p^{-A_4} \,\pi_p(s-1), 
\quad s \in [p], 
\label{eq:prior_cond}
\end{align}

 Several choices of $\pi(s)$ satisfy \eqref{eq:prior_cond}. In particular, a natural sparsity-inducing formulation arises by placing a Beta prior on the overall inclusion probability $w$, then generating each $z_i$ independently from $\mathrm{Bernoulli}(w)$. This leads to the hierarchical formulation:
\begin{align*}
w &\sim \mathrm{Beta}(a_0, b_0), \\
z_i \,\big|\, w &\sim \mathrm{Bernoulli}(w), \\
\theta_i \,\big|\, z_i &\sim z_i\, \mathrm{Lap}(\lambda) + (1-z_i)\, \delta_0, 
\end{align*}
where $a_0$ and $b_0$ are hyperparameters to be specified. 

\cite{ray2022variational} considered both Laplace and Gaussian priors, but we focus on the former due to its superior performance in the KL setting. With the hierarchical Laplace model specified, we now shift our focus to the main ideas driving our VI formulation. Let $\Pi(\theta \mid Y)$ denote the exact posterior distribution on $\theta$. Our goal is to approximate $\Pi(\theta \mid Y)$ with a tractable variational distribution. Traditional mean-field VI approximates an intractable posterior distribution $\Pi$ by solving
\[
\widetilde{\Pi} = \underset{Q \in \mathcal{Q}}{\mathrm{arg\,min}}\ \mathrm{KL}\!\left(Q \,\big\|\, \Pi \right),
\]
where $\mathcal{Q}$ is a tractable family of candidate densities.  
The Kullback–Leibler divergence
\[
\mathrm{KL}(Q \,\|\, \Pi) = \int \log \!\left( \frac{dQ}{d\Pi} \right) \, dQ
\]
measures the discrepancy between $\Pi$ and $Q$. For our choice of $\mathcal{Q}$, we adopt a mean-field family $\mathcal{P}_{\text{MF}}$ consisting of independent mixtures of a Gaussian slab and a Dirac spike at zero:
\begin{align*}
\mathcal{P}_{\text{MF}}
= \left\{ \bigotimes_{i=1}^p \left[ \gamma_i \nm(\mu_i, \sigma_i^2) + (1-\gamma_i)\, \delta_0 \right] : \gamma_i \in [0,1],\ \mu_i \in \RR,\ \sigma_i > 0 \right\}.
\end{align*}
Here, $\mu = (\mu_1,\dots,\mu_p)^\top$, $\sigma^2 = (\sigma_1^2,\dots,\sigma_p^2)^\top$, and $\gamma = (\gamma_1,\dots,\gamma_p)^\top$ are the variational parameters.  In the traditional KL setting, we need to solve 
\begin{align}
\widetilde{\Pi} = \underset{P_{\mu,\sigma,\gamma} \in \mathcal{P}_{\text{MF}}}{\mathrm{arg\,min}}\, \mathrm{KL}\!\left(P_{\mu,\sigma,\gamma}(\theta) \,\big\|\, \Pi(\theta \mid Y)\right).
\label{eq:KLobjective}
\end{align}
One simple hyperparameter choice for the Beta prior for $w$ is to set $a_0$ equal to $ \sum_i \gamma_i$ (the number of non-zero coefficients) and $b_0$ equal to $p - \sum_i \gamma_i$ (the number of zero coefficients).

\subsection{Optimization for variational inference}

We derive two algorithms: a CAVI algorithm and a stochastic VI algorithm. 

A standard and widely used algorithm for solving the optimization problem in \ref{eq:KLobjective} is CAVI, which proceeds by iteratively optimizing each variational parameter (from example $\mu_i$, $\sigma^2_i$, or $\gamma_i$) while holding all others fixed. Each step of the method performs single-variable optimization from an update equation, starting from an initial set of parameter values. The component-wise variational updates are either in closed-form or, if no nice closed-form solution exists, must be obtained via one-variable numerical optimization. 

Stochastic VI is also iterative, but uses stochastic optimization. At each iteration, the method computes the gradient of the ELBO with respect to the variational parameters and performs a gradient step towards the optimum. In this stochastic optimization framework, the ELBO and its gradients can be approximated via Monte Carlo by sampling from the proposed variational family.

\subsection{Rényi divergence variational inference}

While the KL divergence is the standard choice in the VI literature \citep{blei2017variational}, it is only one instance of a broader class of divergences.  
A natural generalization is the Rényi's $\alpha$-divergence \citep{li2016renyi}, defined for $\alpha > 0, \ \alpha \neq 1$ by
\[
D_\alpha(Q \,\|\, P) = \frac{1}{\alpha - 1} \log \int \left( \frac{dQ}{dP} \right)^\alpha \, dP.
\]
As $\alpha \to 1$, $D_\alpha(Q\|P) \to \mathrm{KL}(Q\|P)$, recovering the standard VI formulation. Varying $\alpha$ yields several other well-known divergences, namely, $\alpha = 0.5$ is associated with the Hellinger-related divergence $D_{0.5}(Q \| P) = \displaystyle -2 \log (1- \operatorname{Hel}^2[Q \| P]$ and $\alpha = 2$ the $\chi^2$-related $-\log(1-\chi^2[Q \| P])$. Using $D_\alpha$ in place of KL offers flexibility: different $\alpha$ values yield different behaviors. When $\alpha > 1$, we have a zero-forcing regime where we concentrate on high-density regions of the posterior. When $\alpha <1$, this is the mass-covering regime where we capture more of the posterior's support. See \citep[][]{nowozin2016fgantraininggenerativeneural, poole2016improvedgeneratorobjectivesgans, pmlr-v206-ting-li23a} for applications that explore the mass-covering, zero-forcing, and mode-seeking properties of $f$-divergences. Table~\ref{tab:alpha_behavior} summarizes the different $\alpha$ regimes.

\begin{table}[ht]
\centering
\renewcommand{\arraystretch}{1.15}
\begin{tabular}{ccl}
\toprule
$\alpha$ & Divergence Type & Description \\
\midrule
$\alpha \to 1$ & KL$(Q\|P)$ & Standard VI objective \\
$\alpha > 1$ & Rényi & Zero-forcing \\
$\alpha < 1$ & Rényi & Mass-covering \\
$\alpha \to 0$ & Reverse KL$(P\|Q)$ & Fully mass-covering \\
\bottomrule
\end{tabular}
\caption{Behavior of $D_\alpha$ for different $\alpha$ regimes in Rényi variational inference.}
\label{tab:alpha_behavior}
\end{table}
\vspace{1mm}
 It is important to note the existence of other $\alpha$-divergences in the literature. Among the most common definitions are Amari's and Tsallis'. Amaris's  is defined as:  
\[
D_\alpha(Q \,\|\, P) = \frac{4}{ 1 - \alpha^2} \biggl( 1 - \int \biggl(\frac{dQ}{dP}\biggr)^{\frac{1+\alpha}{2}} \, dP \biggr),
\]
while Tsallis' takes the form of:
\[
D_\alpha(Q \,\|\, P) = \frac{1}{\alpha - 1} \log \int \left( \frac{dQ}{dP} \right)^\alpha \,  dP - 1.
\]
All of these definitions, including Rényi's, can be viewed as special cases of a broader class of $f$-divergences \citep{minka2005divergence}. For our work, we replace the Kullback–Leibler divergence in \eqref{eq:KLobjective} with the Rényi's $\alpha$-divergence:
\begin{align}
    \widetilde{\Pi} = \underset{P_{\mu,\sigma,\gamma} \in \mathcal{P}_{\text{MF}}}{\mathrm{arg\,min}}\, D_{\alpha}\!\left(P_{\mu,\sigma,\gamma}(\theta) \,\big\|\, \Pi(\theta \mid Y)\right). 
\label{eq:objective}
\end{align}


\subsection{Variational Rényi bound}
Minimizing the KL objective function \eqref{eq:KLobjective} in the standard VI setting can be intractable. In the literature, it is common to maximize the evidence lower bound (ELBO). Minimizing KL over the mean-field family $\mathcal{P}_{\mathrm{MF}}$ is equivalent to maximizing
\[
\log p(Y) - \text{KL}(P_{\mu,\sigma,\gamma} \,\|\, p(\theta \mid Y))= \mathbb{E}_{P_{\mu,\sigma,\gamma}} \!\left[ \log \frac{p(\theta,Y)}{P_{\mu,\sigma,\gamma}(\theta)} \right],
\]
since the log marginal likelihood $\log p(Y)$ is independent of the variational parameters.
\cite{li2016renyi} define the Rényi's $\alpha$-divergence analogue of ELBO for optimizing \eqref{eq:objective}. The Rényi divergence between a variational approximation $P_{\mu,\sigma,\gamma}$ and the posterior $p(\theta \mid Y)$ can be written as
\[
D_{\alpha}(P_{\mu,\sigma,\gamma} \,\|\, p(\theta \mid Y))
= \frac{1}{\alpha - 1} \log \mathbb{E}_{P_{\mu, \sigma, \gamma}} \left[ \left(\frac{P_{\mu, \sigma, \gamma}(\theta)}{p(\theta \mid Y)}\right)^{\alpha - 1} \right].
\]

\medskip

Motivated by the ELBO structure, \cite{li2016renyi} define the Variational Rényi (VR) bound as:
\[
\mathcal{L}_{\alpha}(P_{\mu, \sigma, \gamma}) = \log p(Y) - D_{\alpha}\!\left(P_{\mu, \sigma, \gamma} \,\|\, p(\theta \mid Y)\right).
\]
Expanding $\mathcal{L}_\alpha$ in terms of the joint density $p(\theta,Y)$ gives:
\begin{align*}
\log p(Y) - \frac{1}{\alpha-1} \log \int q(\theta)^{\alpha} \, p(\theta \mid Y)^{\,1 - \alpha} \, d\theta 
&= \frac{1}{1-\alpha} \log \int p(\theta, Y)^{\,1-\alpha} \, q(\theta)^{\alpha}  \, d\theta \\
&= \frac{1}{1-\alpha} \log \, \mathbb{E}_{q(\theta)} \!\left[ \left(\frac{p(\theta, Y)}{q(\theta)}\right)^{\,1-\alpha} \right].
\end{align*}

In the ELBO case, the objective is an expectation of a logarithm, which has a convenient form due to linearity of expectations and the nice additive structure of logarithms. By contrast, for $\alpha \neq 1$, the Rényi objective involves a log of an expectation of a potentially nonlinear term, making the optimization problem non-trivial. While the VR bound was not used in our CAVI, it was the objective in our stochastic VI algorithm.

\section{Coordinate Ascent Variational Inference (CAVI)}
\label{S:cavi}
We proceed to derive the component-wise variational updates for our CAVI algorithm \textbf{AlphaVB}. We assume $\alpha > 1$. Since $\frac{1}{\alpha-1} > 0$, 
\begin{align*}
\underset{P_{\mu,\sigma,\gamma} \in \mathcal{P}_{MF}}{\text{arg min}} D_{\alpha}(P_{\mu,\sigma,\gamma} \| p(\theta | Y)) &\equiv \underset{P_{\mu,\sigma,\gamma} \in \mathcal{P}_{MF}}{\text{arg min}} \log \mathbb{E}_{\mu,\sigma,\gamma} \left[ \frac{P_{\mu,\sigma,\gamma}(\theta)^{\alpha-1}}{p(\theta| Y)^{\alpha-1}} \right].
\end{align*}

Following the approach of \cite{ray2022variational}, we examine the Rényi divergence between the variational distribution $P_{\mu,\sigma,\gamma}$ and the posterior $\Pi(\cdot \mid Y)$ for a single coordinate $\theta_i$ given $z_i = 1$. In this case, the variational distribution $P_{\mu_i,\sigma_i \mid z_i=1}$ is a continuous density, while the spike component at $z_i=0$ corresponds to the Dirac measure $\delta_0$. Because the continuous part is singular with respect to $\delta_0$, the Radon--Nikodym derivative reduces to the ratio of the continuous variational density to the continuous prior density. This simplification means we can ignore the spike when updating $\mu_i$ and $\sigma_i$ for $z_i=1$.

Let us fix the latent variable $z_i=1$ and all variational factors except $\mu_i$ or $\sigma_i$ (i.e. using vector notation, ${\mu}_{-i},{\sigma},{\gamma}$ or ${\mu},{\sigma}_{-i},{\gamma}$ are all fixed). Conditioning on $z_i = 1$, we rewrite $\mathbb{E}_{\mu,\sigma,\gamma | z_i = 1} \left[ \frac{P_{\mu,\sigma,\gamma}(\theta)^{\alpha-1}}{p(\theta| Y)^{\alpha-1}} \right]$ as a function of $\mu_i$ or $\sigma_i$. 
 The Laplace density centered at $0$ is given by
$f_\lambda(\theta_i) = \frac{\lambda}{2} e^{-\lambda |\theta_i|}$. The prior inclusion probability equals:
\begin{align*}
    \mathbb{P}(z_i = 1) &= \int_{0}^1 \PP(z_i = 1, w) dw
    = \int_{0}^1 w \cdot \text{Beta}(a_0, b_0) dw 
    =  \mathbb{E}[w] = \frac{a_0}{a_0 + b_0}.
\end{align*}

 Let $\bar{w}_i = \frac{a_0}{a_0 + b_0}$. We evaluate the Radon–Nikodym derivative and aim to maximize
\begin{align*}
\mathbb{E}_{\mu, \sigma, \gamma | z_i = 1} \left[ \left( \frac{\frac{d P_{\mu_{-i}, \sigma_{-i}, \gamma_{-i} | z_i = 1} \otimes \nm(\mu_i, \sigma_i^2)}{d\pi_{-i} \otimes \bar{w_i}Lap(\lambda)}}{D_{\pi}^{-1} \exp \left\{ \frac{1}{2} \left( -\frac{\|Y - X\theta\|_2^2}{2} \right) \right\}} \right)^{\alpha-1}\right] 
\end{align*} which is equivalent to maximizing
\begin{equation}
\scalebox{0.8}{%
$\mathbb{E}_{\mu, \sigma, \gamma | z_i = 1} \!\left[\exp\left\{(\alpha - 1)\!\left(
- (Y^\top X)_i \theta_i 
+ \frac{1}{2} \theta_i^2 (X^\top X)_{ii} +  \theta_i \sum_{j\neq i} \theta_j (X^\top X)_{ji} 
+ \lambda |\theta_i| 
- \frac{(\theta_i - \mu_i)^2}{2\sigma_i^2} 
- \log \sigma_i
\right)\right\}\right]$%
}
\end{equation}


\noindent Evaluating the expectation is intractable and there is no straightforward closed-form solution to the best of our knowledge.  We apply the multivariate delta method \citep{braun2008}, with details of these derivations left to Appendix~\ref{A:delta}. Applying the delta method, we have
\[
\mathbb{E}[f(\theta)] 
\approx f(\mathbb{E}[\theta]) 
+ \frac{1}{2} \operatorname{tr} \bigl(\nabla^2 f(\mathbb{E}[\theta]) \cdot \operatorname{Cov}(\theta)\bigr).
\]

\noindent Evaluating the function $g$ at $\mathbb{E}[\theta]$, we find

\begin{align*}
g(\mathbb{E}[\theta]) 
&= \exp\biggl\{(\alpha - 1)\biggl(
- (Y^\top X)_i \mu_i 
+ \frac{1}{2} \mu_i^2 (X^\top X)_{ii} 
+ \mu_i \sum_{j \neq i} \gamma_j \mu_j (X^\top X)_{ji} \\
&+ \lambda \sqrt{\mu_i^2 + \varepsilon}
- \frac{(\mu_i - \mu_i)^2}{2\sigma_i^2} 
- \log \sigma_i
\biggr)\biggr\}.
\end{align*}

\noindent This simplifies to
\[
g(\mathbb{E}[\theta]) = \exp\!\left\{(\alpha - 1)\!\left(
- (Y^\top X)_i \mu_i 
+ \frac{1}{2} \mu_i^2 (X^\top X)_{ii} 
+ \mu_i \sum_{j \neq i} \gamma_j \mu_j (X^\top X)_{ji} 
+ \lambda \sqrt{\mu_i^2 + \varepsilon}
- \log \sigma_i
\right)\right\}.
\]
 Focusing on the term
$
\frac{1}{2} \operatorname{tr} \bigl(\nabla^2 g(\mathbb{E}[\theta]) \cdot \operatorname{Cov}(\theta)\bigr),
$
since the covariance matrix $\operatorname{Cov}(\theta)$ is diagonal, this expression equals
\[
\frac{1}{2} \sum_{k=1}^p 
\frac{\partial^2 g}{\partial \theta_k^2}(\mathbb{E}[\theta]) \, \left[\operatorname{Cov}(\theta)\right]_{kk}.
\]

\noindent Taking the second partial derivatives of $g$ and evaluating at $\mathbb{E}[\theta]$, we have
\begin{align*}
 \frac{\partial^2 g}{\partial \theta_i^2} \bigl(\mathbb{E}[\theta]\bigr)
&= g\!\bigl(\mathbb{E}[\theta]\bigr) \cdot(\alpha - 1)^2 \left(- (Y^\top X)_i + \mu_i (X^\top X)_{ii} + \sum_{j \neq i} (X^\top X)_{ji} \gamma_j \mu_j + \lambda (\mu_i^2 + \varepsilon)^{-\frac{1}{2}} \mu_i
\right)^2  \\ 
&+ g\!\bigl(\mathbb{E}[\theta]\bigr) \cdot (\alpha -1) \left((X^\top X)_{ii} - \tfrac{1}{\sigma_i^2} + \lambda \cdot ((\mu_i^2 + \varepsilon)^{-\frac{1}{2}}  - \mu_i^2 (\mu_i^2 + \varepsilon)^{-\frac{3}{2}}) \right),   
\end{align*}
 and if $k \neq i$
\begin{align*}
\frac{\partial^2 g}{\partial \theta_k^2} \bigl(\mathbb{E}[\theta]\bigr)
&= (\alpha - 1)^2 \bigl( \mu_i (X^\top X)_{ki} \bigr)^2 g\!\bigl(\mathbb{E}[\theta]\bigr).
\end{align*}
\noindent Putting it all together, let us define
\[
\kappa_i(\mu_i, \sigma_i) = g(\mathbb{E}[\theta]) \cdot \left(1 + \frac{(\alpha - 1)^2}{2} A_i(\mu_i, \sigma_i) + \frac{\alpha - 1}{2} B_i(\mu_i, \sigma_i) + \frac{(\alpha - 1)^2}{2}C_i(\mu_i) \right),
\]
where
\begin{align*}
A_i(\mu_i, \sigma_i) 
&= \left(
- (Y^\top X)_i + \mu_i (X^\top X)_{ii} 
+  \sum_{j \neq i} \gamma_j \mu_j (X^\top X)_{ji} 
+ \lambda (\mu_i^2 + \varepsilon)^{-\frac{1}{2}} \mu_i
\right)^2 \sigma_i^2, \\
B_i(\mu_i, \sigma_i) 
&=  (X^\top X)_{ii} \sigma_i^2 - 1 + \lambda \cdot \sigma_i^2 \cdot \left((\mu_i^2 + \varepsilon)^{-\frac{1}{2}}  - \mu_i^2 (\mu_i^2 + \varepsilon)^{-\frac{3}{2}}\right), \\
C_i(\mu_i) &= \mu_i^2 \sum_{\substack{k=1 \\ k \neq i}}^{p} \left[
\left( (X^\top X)_{ki} \right)^2 
\left( \gamma_k(1 - \gamma_k) \mu_k^2 + \gamma_k \sigma_k^2 \right)
\right].
\end{align*}

\noindent Hence, the coordinate update for \(\mu_i\) or \(\sigma_i\) reduces to:
\[
\underset{\mu_i, \sigma_i}{\arg\max} \ \log \kappa_i(\mu_i, \sigma_i) .
\]

\noindent The objective function is:

\begin{equation*}
    \begin{aligned}
\log \kappa_i &= (\alpha - 1)\!\left(
- (Y^\top X)_i \mu_i 
+ \frac{1}{2} \mu_i^2 (X^\top X)_{ii} 
+ \mu_i \sum_{j \neq i} \gamma_j \mu_j (X^\top X)_{ji} 
+ \lambda \sqrt{\mu_i^2 + \varepsilon}
- \log \sigma_i
\right) \nonumber \\
&\quad \quad + \log \left(1 + \frac{(\alpha - 1)^2}{2} A_i(\mu_i, \sigma_i) + \frac{\alpha -1}{2} B_i(\mu_i, \sigma_i) + \frac{(\alpha - 1)^2}{2}C_i(\mu_i) \right).
\end{aligned}
\end{equation*}

\noindent This leads to simpler update equations:

\begin{equation}
\scalebox{0.9}{$
\begin{aligned}
\log \kappa_i(\mu_i)
&= (\alpha - 1)\!\left(
- (Y^\top X)_i \mu_i
+ \frac{1}{2} \mu_i^2 (X^\top X)_{ii}
+ \mu_i \sum_{j \neq i} \gamma_j \mu_j (X^\top X)_{ji}
+ \lambda \sqrt{\mu_i^2 + \varepsilon}
\right) \\
&\quad + \log \Biggl(
1 + \frac{(\alpha - 1)^2}{2} A_i(\mu_i, \sigma_i)
+ \frac{\alpha - 1}{2} B_i(\mu_i, \sigma_i)
+ \frac{(\alpha - 1)^2}{2} C_i(\mu_i)
\Biggr)
\end{aligned}
$}
\label{eq:cavi-mu}
\end{equation}

\noindent and 

\begin{equation}
\scalebox{0.85}{$
\begin{aligned}
\log \kappa_i(\sigma_i)
&= -(\alpha -1) \log \sigma_i
+ \log \biggl(
1 + \frac{(\alpha - 1)^2}{2} A_i(\mu_i, \sigma_i)
+ \frac{\alpha-1}{2} B_i(\mu_i, \sigma_i)
+ \frac{(\alpha - 1)^2}{2} C_i(\mu_i)
\biggr)
\end{aligned}
$}
\label{eq:cavi-sigma}
\end{equation}

\noindent The update equation for $\gamma_i$ is a little more challenging. We aim to maximize

\begin{equation*}
C \cdot \mathbb{E}_{\mu, \sigma, \gamma} \left[ \left( \frac{\frac{d P_{\mu_{-i}, \sigma_{-i}, \gamma_{-i} }}{d\pi_{-i}}(\theta_{-i}) \cdot \frac{d(\gamma_i \nm(\mu_i, \sigma_i^2) + (1-\gamma_i)\delta_0)}{d(\bar{w}_i \text{Lap}(\lambda) + (1-\bar{w}_i)\delta_0}(\theta_i)}{ \exp \left\{ -\frac{\|Y - X\theta\|_2^2}{2}  \right\}} \right)^{(\alpha - 1)} \right],
\end{equation*} 

\noindent where $C>0$ is independent of $\gamma_i$ and $\bar{w}_i = \frac{a_0}{a_0+b_0}$. Define $q_i = \gamma_i \cdot \mathcal{N}(\mu_i, \sigma_i^2) + (1 - \gamma_i) \cdot \delta_0 $ and $ r_i = \bar{w}_i \cdot \text{Lap}(\lambda) + (1 - \bar{w}_i) \cdot \delta_0.$ The Radon–Nikodym derivative of \(q_i\) with respect to \(r_i\) is given by

\begin{equation*}
\frac{dq_i}{dr_i}(\theta_i) =
\begin{cases}
\displaystyle \frac{\gamma_i \cdot d\nm(\mu_i, \sigma_i^2)}{\bar{w}_i \cdot d\text{Lap}(\lambda)} & \text{if } z_i = 1 \\[10pt]
\displaystyle \frac{1 - \gamma_i}{1 - \bar{w}_i} & \text{if } z_i = 0
\end{cases}
\end{equation*}

\noindent We are left with a nicer looking expression:

\begin{equation*}
\mathbb{E}_{\mu, \sigma, \gamma} \left[ \left(\mathbb{I}_{\{z_i = 1\}} \cdot \frac{\gamma_i d\nm(\mu_i, \sigma_i^2)}{\bar{w}_id \text{Lap}(\lambda)}(\theta_i) + \mathbb{I}_{\{z_i = 0\}} \cdot \frac{1-\gamma_i}{1-\bar{w}_i} \right)^{(\alpha - 1)} \exp \left\{ \frac{(\alpha - 1)}{2} \|Y - X\theta\|_2^2 \right\} \right].
\end{equation*}

\noindent Observe that 

\begin{equation*}
\begin{aligned}
       \mathbb{I}_{\{z_i = 1\}} \cdot \frac{\gamma_i d\nm(\mu_i, \sigma_i^2)}{\bar{w}_id \text{Lap}(\lambda)}(\theta_i) + \mathbb{I}_{\{z_i = 0\}} \cdot \frac{1-\gamma_i}{1-\bar{w}_i} = \exp &\Biggl( \mathbb{I}_{\{z_i = 1\}} \cdot \log \frac{\gamma_i d\nm(\mu_i, \sigma_i^2)}{\bar{w}_id \text{Lap}(\lambda)}(\theta_i) \\
       &\quad + \mathbb{I}_{\{z_i = 0\}} \cdot \log \frac{1-\gamma_i}{1-\bar{w}_i} \Biggr).
\end{aligned}
\end{equation*}

\noindent Our final display nicely reduces to

\begin{equation*}
\begin{aligned}
         \mathbb{E}_{\mu, \sigma, \gamma} \Biggl[ \exp \Biggl\{ (\alpha - 1) &\Biggl(\frac{1}{2} (Y - X\theta)^\top (Y - X\theta) +  \mathbb{I}_{\{z_i = 1\}} \cdot \log \frac{\gamma_i d\nm(\mu_i, \sigma_i^2)}{\bar{w}_id \text{Lap}(\lambda)}(\theta_i) \\
         &\quad+ \mathbb{I}_{\{z_i = 0\}} \cdot \log \frac{1-\gamma_i}{1-\bar{w}_i} \Biggr)  \Biggr\} \Biggr].
\end{aligned}
\end{equation*}

\noindent When maximizing the last display, we treat $\theta_{-i}$ as fixed and remove extraneous terms from $||Y - X\theta||^2_2$, as was done previously. We also have that:
\[
    \frac{d \nm(\mu_i, \sigma_i^2)}{d \text{Lap}(\lambda)} (\theta_i) 
    = \frac{\sqrt{2}}{\sqrt{\pi} \sigma_i \lambda} 
    \exp\left( -\frac{(\theta_i - \mu_i)^2}{2\sigma_i^2} + \lambda |\theta_i| \right).
\]

\noindent Finally, we maximize the following expression:
\begin{equation*}
\begin{aligned}
\mathbb{E}_{\mu, \sigma, \gamma} \Bigg[ 
& \exp \Bigg\{ (\alpha - 1) \Bigg(
- (Y^\top X)_i \theta_i 
+ \tfrac{1}{2} \theta_i^2 (X^\top X)_{ii} 
+  \theta_i \sum_{j \neq i} (X^\top X)_{ji} \theta_j \\
& \quad + \mathbb{I}_{\{z_i = 1\}} \Bigg( 
\log \frac{\sqrt{2}}{\sqrt{\pi} \, \sigma_i \lambda} 
- \frac{(\theta_i - \mu_i)^2}{2 \sigma_i^2} 
+ \lambda |\theta_i| 
+ \log \frac{\gamma_i}{\bar{w}_i}
\Bigg) + \mathbb{I}_{\{z_i = 0\}} 
\log \frac{1 - \gamma_i}{1 - \bar{w}_i}
\Bigg) \Bigg\} 
\Bigg].
\end{aligned}
\end{equation*}

Here, instead of applying the delta method, we use Jensen's inequality; see Appendix~\ref{A:jensen} for details. Putting everything together, we recover the exact same update as in \cite{ray2022variational}:

\begin{equation}
 \begin{aligned}
h_i(\gamma_i|\boldsymbol{\mu},\boldsymbol{\sigma},\boldsymbol{\gamma}_{-i}) = \gamma_i \bigg\{ \mu_i \sum_{j\neq i} (X^TX)_{ji} \gamma_j\mu_j+\tfrac{1}{2} (X^TX)_{ii}(\sigma_i^2+\mu_i^2) -(Y^TX)_i\mu_i +\log \frac{\sqrt{2}}{\sqrt{\pi}\sigma_i\lambda} -\frac{1}{2} \nonumber\\
+ \lambda \sigma_i\sqrt{2/\pi}e^{-\mu_i^2/(2\sigma_i^2)} +\lambda\mu_i(1-2\Phi(-\mu_i/\sigma_i)) + \log \frac{\gamma_i}{1-\gamma_i} + \log \frac{b_0}{a_0} \bigg\} + \log(1-\gamma_i) + C.\nonumber\\
\end{aligned}
\end{equation}

\noindent where $C > 0$ is independent of $\gamma_i$. Taking the derivative of $h_i$ with respect to $\gamma_i$ and setting it equal to zero solves:

\begin{equation}
\begin{aligned}
\log \frac{\gamma_i}{1-\gamma_i}&=\log \frac{a_0}{b_0}+\log \frac{\sqrt{\pi}\sigma_i \lambda}{\sqrt{2}}+
(Y^TX)_i \mu_i- \mu_i \sum_{k\neq i} (X^TX)_{ik}\gamma_k\mu_k-\tfrac{1}{2}(X^TX)_{ii}(\sigma_i^2 + \mu_i^2) \nonumber\\
&\qquad- \lambda\sigma_i\sqrt{2/\pi}e^{-\mu_i^2/(2\sigma_i^2)}- \lambda\mu_i(1-2\Phi(-\mu_i/\sigma_i))+\frac{1}{2}=:\Gamma_i(\boldsymbol{\mu},\boldsymbol{\sigma},\boldsymbol{\gamma}_{-i})
\label{gamma-update}
\end{aligned}
\end{equation}

\noindent The update equation for $\gamma_i$ given $\mu, \sigma, \gamma_{-i}$ is $\gamma_i=\text{logit}^{-1}\big(\Gamma_i(\boldsymbol{\mu},\boldsymbol{\sigma},\boldsymbol{\gamma}_{-i}) \big)$.

We present the finalized CAVI with Laplace spike and slab prior. The binary entropy function $H: [0,1] \to \mathbb{R}$ is defined as $H(z) = -z\log_2(z)- (1-z)\log_2(1-z)$. We employ a prioritized order scheme for $a$ where we using MLE estimation for $\theta$ and order by the highest to lowest.

\begin{algorithm}[H]
\caption{AlphaVB CAVI with Laplace Spike-and-Slab Prior}
\begin{algorithmic}[1]
\State \textbf{Initialize}: $(\Delta_H,\sigma,\gamma)$, $\mu = \hat{\mu}^{(0)}$ (for a preliminary estimator $\hat{\mu}^{(0)}$), $a = \text{order}(|\mu|)$
\While{$\Delta_H \geq \varepsilon$}
  \For{$j = 1$ to $p$}
    \State $i = a_j$
    \State $\mu_i = \displaystyle\arg\min_{\mu_i} \log \kappa_i(\mu_i \mid \mu_{-i}, \sigma, \gamma, z_i = 1)$ $\eqref{eq:cavi-mu}$
    \State $\sigma_i = \displaystyle\arg\min_{\sigma_i} \log \kappa_i(\sigma_i \mid \mu, \sigma_{-i}, \gamma, z_i = 1)$ $\eqref{eq:cavi-sigma}$
    \State $\gamma_{\text{old},i} = \gamma_i$, \quad $\gamma_i = \text{logit}^{-1}\big(\Gamma_i(\boldsymbol{\mu},\boldsymbol{\sigma},\boldsymbol{\gamma}_{-i}) \big)$ \eqref{gamma-update}
  \EndFor
  \State $\Delta_H = \max_i \left\{ \left| H(\gamma_i) - H(\gamma_{\text{old},i}) \right| \right\}$
\EndWhile
\end{algorithmic}
\end{algorithm}

\section{Stochastic Variational Inference}
\label{S:mc_vb}
After deriving the variational updates for our AlphaVB CAVI method in Section \ref{S:cavi}, we now return to the Rényi variational lower bound introduced in Section \ref{S:background}. Our goal in this section is to derive a stochastic optimization method, which we call AlphaSVB, by replacing the expectations in the Rényi bound with a Monte Carlo estimate and then computing the gradient of this approximation with respect to the variational parameters. This stochastic approach is an alternative to the deterministic updates produced by CAVI. 

\subsection{Monte Carlo (MC) Approximation of VR Bound}

Consider the joint model $p(\theta, z, Y | w)$. We set $w = \frac{a_0}{a_0 + b_0}$ and we drop it from the joint for notational convenience to obtain:
\begin{align*}
p(\theta, z, Y) &= p(Y | \theta, z) \, p(\theta, z) = p(Y | \theta, z) \, p(\theta | z) \, p(z) \\
&= \nm(Y; X\theta, \sigma^2 I) \prod_{i=1}^p \left[ z_i \, \text{Lap}(\theta_i; \lambda) + (1-z_i) \, \delta_0(\theta_i) \right] \cdot \prod_{i=1}^p \text{Bern}(z_i; w) \\
&= \nm(Y; X\theta, \sigma^2 I) \prod_{i=1}^p \left[ z_i \, \text{Lap}(\theta_i; \lambda) + (1-z_i) \, \delta_0(\theta_i) \right] \cdot \prod_{i=1}^p w^{z_i} (1-w)^{1-z_i}.
\end{align*}
 Recall that the Rényi variational lower bound is:
\begin{align*}
\mathcal{L}_\alpha(P_{\mu,\sigma,\gamma}; Y)
&= \frac{1}{1-\alpha} \log \mathbb{E}_{\mu,\sigma,\gamma} \left[ 
\frac{p(\theta, z, Y)^{1-\alpha}}{P_{\mu,\sigma,\gamma}(\theta)^{1-\alpha}} 
\right].
\end{align*}

The Monte Carlo (MC) approximation of the above using $K$ samples is:
\begin{align*}
\widehat{\mathcal{L}}_\alpha(P_{\mu,\sigma,\gamma}; Y) 
&= \frac{1}{1-\alpha} 
\log \left[ \frac{1}{K} \sum_{j=1}^K 
\left( 
\frac{p(\theta^{(j)}, z^{(j)}, Y)}
{P_{\mu,\sigma,\gamma}(\theta^{(j)})}
\right)^{1-\alpha} \right],
\end{align*}
where $(\theta^{(j)}, z^{(j)})$ are sampled from the variational family.
\subsection{Sampling}
Here is an overview of the sampling procedure used in our stochastic VI method. At iteration $k$, given the current variational parameters $(\mu^{(k-1)}, \sigma^{(k-1)}, \gamma^{(k-1)})$, we generate $K$ samples as follows:  
\begin{enumerate}
    \item For each $j\in [K]$:
    \begin{enumerate}
        \item For every $i \in [p]$, sample $
        z_i^{(j)} \sim \text{Bern}\big(\gamma_i^{(k-1)}\big)$.
        \item For every $i \in [p]$, conditional on $z_i^{(j)}$:
        \begin{itemize}
            \item If $z_i^{(j)} = 0$, set $\theta_i^{(j)} = 0$ (inactive signal).
            \item If $z_i^{(j)} = 1$, sample the active signal:
            $\theta_i^{(j)} \sim \nm\big(\mu_i^{(k-1)}, \big(\sigma_i^{(k-1)}\big)^2\big)$.
        \end{itemize}
    \end{enumerate}
\end{enumerate}
This sampling scheme provides $(\theta^{(j)}, z^{(j)})$ pairs that are then used to compute gradients.

\subsection{Unified Framework}

Following \cite{li2016renyi}, the gradient of the Monte Carlo approximation of the Rényi variational bound is given by:
\[
\displaystyle
\nabla_{\mu,\sigma,\gamma} \, \widehat{\mathcal{L}}_\alpha(P_{\mu,\sigma,\gamma}; Y)
= \frac{1}{K} \sum_{j=1}^K \widehat{w}_\alpha^{(j)} \,
\nabla_{\mu,\sigma,\gamma} \log 
\frac{p(\theta^{(j)}, z^{(j)}, Y)}
{P_{\mu,\sigma,\gamma}(\theta^{(j)})},
\]
where the importance weights are defined as:
\[
\widehat{w}_\alpha^{(j)}
= \frac{\left[ \dfrac{p(\theta^{(j)}, z^{(j)}, Y)}{P_{\mu,\sigma,\gamma}(\theta^{(j)})} \right]^{1-\alpha}}
{\displaystyle \sum_{\ell=1}^K 
\left[ \dfrac{p(\theta^{(\ell)}, z^{(\ell)}, Y)}{P_{\mu,\sigma,\gamma}(\theta^{(\ell)})} \right]^{1-\alpha}}.
\]

\noindent The variational parameters are updated via a gradient ascent (for maximization) step:
\[
(\mu^{(k)}, \sigma^{(k)}, \gamma^{(k)})
= (\mu^{(k-1)}, \sigma^{(k-1)}, \gamma^{(k-1)})
+ \eta \, \nabla_{\mu,\sigma,\gamma} \widehat{\mathcal{L}}_\alpha,
\]
where \( \eta > 0 \) is the learning rate. Consider the log-ratio term:
\begin{align*}
\log 
\frac{p(\theta, z, Y)}
{P_{\mu,\sigma,\gamma}(\theta)} 
&= \log p(\theta, z, Y) - \log P_{\mu,\sigma,\gamma}(\theta) \\
&= \log \nm(Y; X\theta, \sigma^2 I) 
+ \sum_{i=1}^p \log \left[ z_i \, \text{Lap}(\theta_i; \lambda) + (1-z_i) \, \delta_0(\theta_i) \right] \\ 
&\quad + \sum_{i=1}^p \left[ z_i \log w + (1-z_i) \log (1-w) \right] \\
&\quad - \sum_{i=1}^p \log \left[ \gamma_i \, \nm(\theta_i \mid \mu_i, \sigma_i^2) + (1-\gamma_i) \, \delta_0(\theta_i) \right].
\end{align*}

\noindent Then,
\begin{align*}
\log 
\frac{p(\theta, z, Y)}
{P_{\mu,\sigma,\gamma}(\theta)} 
&=  \frac{-n}{2} (\log(2\pi) + \log \sigma^2 ) - \frac{1}{2\sigma^2} \|Y-X\theta\|^2
+ \sum_{i=1}^p z_i \log  \text{Lap}(\theta_i; \lambda)  \\ 
&\quad + \sum_{i=1}^p \left[ z_i \log w + (1-z_i) \log (1-w) \right]  \\
&\quad - \sum_{i=1}^p \log \left[ P_{\mu, \sigma, \gamma}(\theta_i ) \right].
\end{align*}

By linearity of the gradient operator, the gradient of this log-ratio with respect to the variational parameters becomes:
\[
\nabla_{\mu, \sigma, \gamma} 
\log \frac{p(\theta, z, Y)}{P_{\mu,\sigma,\gamma}(\theta)} 
= - \sum_{i=1}^p \nabla_{\mu, \sigma, \gamma} 
\log \left[ P_{\mu, \sigma, \gamma}(\theta_i ) \right].
\]

We now consider the coordinate-wise gradients of the log variational density, explicitly conditioning on \( z_i \). Recall the variational density is:
\[
P_{\mu, \sigma, \gamma \mid z_i}(\theta_i ) = 
\begin{cases}
(1 - \gamma_i) \cdot \delta_0(\theta_i), & \text{if } z_i = 0, \\
\gamma_i \cdot \nm(\theta_i \mid \mu_i, \sigma_i^2), & \text{if } z_i = 1.
\end{cases}
\]

\noindent If \( z_i = 1 \), then \( \theta_i \sim \nm(\mu_i, \sigma_i^2) \), and the variational density becomes:
  \[
  \log P_{\mu, \sigma, \gamma \mid z_i = 1}(\theta_i )= \log \gamma_i + \log \nm(\theta_i \mid \mu_i, \sigma_i^2).
  \]
  We compute:
  \begin{align*}
  \frac{\partial}{\partial \mu_i} \log P_{\mu, \sigma, \gamma \mid z_i = 1}(\theta_i )
  &= \frac{\partial}{\partial \mu_i} \log \nm(\theta_i \mid \mu_i, \sigma_i^2)
  =  \frac{\partial}{\partial \mu_i} ( \frac{-(\theta_i - \mu_i)^2}{2\sigma_i^2})= \frac{\theta_i - \mu_i}{\sigma_i^2} , \\
  \frac{\partial}{\partial \sigma_i} \log P_{\mu, \sigma, \gamma \mid z_i = 1}(\theta_i ) 
  &= \frac{\partial}{\partial \sigma_i} \log \nm(\theta_i \mid \mu_i, \sigma_i^2)
  = \frac{(\theta_i - \mu_i)^2 - \sigma_i^2}{\sigma_i^3}.
  \end{align*}

 \noindent For the gradient with respect to \( \gamma_i \), regardless of \( z_i \), we use:
  \[
  \log P_{\mu, \sigma, \gamma \mid z_i}(\theta_i ) = z_i \log \gamma_i + (1 - z_i) \log (1 - \gamma_i),
  \]
  and so:
  \[
  \frac{\partial}{\partial \gamma_i} \log P_{\mu, \sigma, \gamma \mid z_i}(\theta_i ) 
  = \frac{z_i}{\gamma_i} - \frac{1 - z_i}{1 - \gamma_i}.
  \]

\noindent Combining these, we obtain these expressions:
\begin{align*}
\frac{\partial}{\partial \mu_i} \log P_{\mu, \sigma, \gamma \mid z_i}(\theta_i ) &= z_i \cdot \frac{\theta_i - \mu_i}{\sigma_i^2}, \\
\frac{\partial}{\partial \sigma_i} \log P_{\mu, \sigma, \gamma \mid z_i}(\theta_i ) &= z_i \cdot \frac{(\theta_i - \mu_i)^2 - \sigma_i^2}{\sigma_i^3}, \\
\frac{\partial}{\partial \gamma_i} \log P_{\mu, \sigma, \gamma \mid z_i}(\theta_i ) &= \frac{z_i}{\gamma_i} - \frac{1 - z_i}{1 - \gamma_i}.
\end{align*}
 Consider the following reparameterization of $\gamma$ to ensure it is a valid probability: $\gamma_i = \text{logit}^{-1} (\tau_i)$, where $\tau_i \in \RR$. Equivalently, $\tau_i = \log \frac{\gamma_i}{1-\gamma_i}$. We also make the reparametrization exp(log) for $\sigma$. We reparameterize the standard deviations by setting 
\[
\sigma_i = \exp(\log \eta_i), \quad \eta_i \in \mathbb{R},
\]
which ensures $\sigma_i > 0$ and unconstrained learning over $\eta_i$.

\subsection{AlphaSVB Algorithm}
 We now present the stochastic VI algorithm for optimizing the Monte Carlo approximation of the Rényi variational bound. At each iteration, we generate samples $(\theta^{(j)}, z^{(j)})$ from the variational distribution, compute normalized importance weights $\widehat{w}_\alpha^{(j)}$, evaluate gradients of the bound, and update the variational parameters $(\mu,\sigma,\gamma)$ using stochastic gradient ascent.
\begin{algorithm}[H]
\caption{AlphaSVB for Rényi Bound with Laplace Spike-and-Slab Prior}
\begin{algorithmic}[1]
\State \textbf{Input:} $Y$, $X$, $p(\theta,z,Y)$, learning rates $(\eta_\mu,\eta_\sigma,\eta_\gamma)$, samples $K$, max iterations $T$
\State \textbf{Init:} $\mu^{(0)}, \sigma^{(0)}, \gamma^{(0)}$
\For{$k=1$ to $T$}
  \State $\nabla_\mu,\nabla_\sigma,\nabla_\gamma \gets 0$
  \For{$j=1$ to $K$}
    \For{$i=1$ to $p$}
      \State $z_i^{(j)} \sim \mathrm{Bern}(\gamma_i^{(k-1)})$
      \State $\theta_i^{(j)} \gets \begin{cases}
          0 & z_i^{(j)}=0 \\
          \mathcal{N}(\mu_i^{(k-1)}, (\sigma_i^{(k-1)})^2) & z_i^{(j)}=1
      \end{cases}$
    \EndFor
    \State Compute $\widehat{w}_{\alpha}^{(j)}$
    \State Accumulate $(\nabla_\mu,\nabla_\sigma,\nabla_\gamma)$
  \EndFor
  \State $\mu^{(k)} \gets \mu^{(k-1)} + \eta_\mu \nabla_\mu$
  \State $\sigma^{(k)} \gets \sigma^{(k-1)} + \eta_\sigma \nabla_\sigma$
  \State $\gamma^{(k)} \gets \gamma^{(k-1)} + \eta_\gamma \nabla_\gamma$
\EndFor
\State \textbf{Output:} $(\mu,\sigma,\gamma)$
\end{algorithmic}
\end{algorithm}


\section{Simulation Study}
\label{S:simulations}
\subsection{Simulation Configurations}

To validate the performance of our proposed AlphaVB and the stochastic AlphaSVB, we conduct a comprehensive simulation benchmark study. All simulation datasets are generated with the simple model in \Cref{eq:data_model}, where $Z$ consists of IID Gaussian noise and $\theta$ is partitioned randomly into signals and non-signals. All true signals' coefficients, as denoted by $\theta_s$, are generated uniformly via $U(-3,3)$, whereas non-signals are 0 by design. For all simulation configurations, we repeat the procedure 100 times to ensure reproducibility.

Let $n$ denote the number of observations, $p$ the number of features in regression, and $s$ the number of true signals. We specifically tested four configurations as summarized in \Cref{tab:sim_configs}. \textbf{Configuration (i)} corresponds to a moderate dimension and sparsity with small sample size, which serves as a baseline for the sparse regression setting; \textbf{configuration (ii)} consists of extra-high-dimensional with 1000 features; \textbf{configuration (iii)} corresponds to that of extreme sparsity with only $6.25\%$ of features as signals; \textbf{configuration (iv)} has relatively large sample size as compared to the number of features.

By default, we chose $\alpha=1.01$ for AlphaVB, and $\alpha =0.9$ for AlphaSVB. A more detailed discussion of the choice of $\alpha$ and our rationale is given in \ref{S:results_alpha}.

\begin{table}[h]
\centering
\renewcommand{\arraystretch}{1.2}
\begin{tabular}{ccccc}
\toprule
Configuration & Purpose & $n$ & $p$ & $s$  \\
\midrule
(i)   & Baseline & 100 & 200 & 10  \\
(ii)  & High Dimension & 400 & 1000 & 40   \\
(iii) & High Sparsity & 200 & 800 & 5  \\
(iv)  & Large Sample & 300 & 450 & 20  \\
\bottomrule
\end{tabular}
\caption{Sparse high-dimensional regression configurations.}
\label{tab:sim_configs}
\end{table}

\subsection{Sparse High Dimensional Regression Benchmark}

To assess the performance of AlphaVB and AlphaSVB, we conduct a benchmark against state-of-the-art Bayesian and frequentist methods for the sparse high-dimensional regression task. We specifically included the following state-of-the-art methods in the field for comparison:

\begin{enumerate}
    \item lasso \citep{tibshirani1996regression}: A frequentist variable selection method that uses penalized linear regression and $k$-fold cross validation from the popular R package \emph{glmnet}.
    \item sparsevb \citep{ray2022variational}: A variational variable selection method using KL divergence and model selection priors that motivated our work.
    \item varbvs \citep{carbonetto2012}: Another variational variable selection method that computes a posterior inclusion probability (PIP) for each variable. 
    \item spikeslab \citep{ishwaran2005, ishwaran2014}: A rescaled spike-and-slab method using the generalized elastic net for coefficient estimation from the \emph{spikeslab} R package.
\end{enumerate}

For benchmarking, we implemented each method using the model choices outlined below. Because the methods return different forms of coefficient and variable selection estimation, we handled and mapped their outputs differently. For lasso, our coefficients $\mu$ were taken from the fitted model corresponding to the regularization hyperparameter (\texttt{lambda.min}) with the smallest Mean Squared Prediction Error (MSPE), excluding the intercept. Since lasso does not provide posterior inclusion probabilities, our $\gamma$ was computed by converting $\mu$ into a binary vector based on whether a variable was selected as significant or not (if $\mu_i = 0$, then $\gamma_i = 0$; otherwise $\gamma_i = 1$). This choice is consistent with lasso shrinking insignificant coefficients to zero \cite{tibshirani1996regression}. For sparsevb, we used a linear-Gaussian model with a Laplace slab prior and set the prior scale parameter to $1$. We initialized the variational parameter $\sigma$ to the $p$-vector of ones. 

For varbvs, after specifying a Gaussian response through \texttt{family == Gaussian}, we set the $n \times p$ covariate data matrix $Z$ as \texttt{NULL}. The coefficient estimates $\mu$ were  obtained from the fitted model's \texttt{beta} output, while $\gamma$ was computed from \texttt{pip}, which is a weighted average of
estimates of posterior inclusion probabilities \cite{carbonetto2012}. Finally, for spikeslab, we set \texttt{bigp.smalln == TRUE} since our setting is high-dimensional ($p >>n$). We took $\mu$ to be the \texttt{gnet.scale} output, which is the generalized elastic net estimator for variable selection \cite{ishwaran2014}, and $\gamma$ was computed according to which estimate was nonzero:
\[\gamma_j = \mathbb{I}(|\mu_j|>0).\]

As a standard suite of benchmark metrics, we used the following criteria: $\ell_2$ Error for regression coefficients $\theta$, False Discovery Rate (FDR) and True Positive Rate (TPR) for sparsity discovery and inference, and Mean Squared Prediction Error (MSPE). We then repeated each simulation 100 times for each method, and the standard deviation for each metric is included for quantification of run-to-run variance.

As shown in \Cref{tab:performance_benchmark}, AlphaVB achieves competitive performance against state-of-the-art methods, whereas AlphaSVB lags behind. Specifically, AlphaVB excels in achieving low FDR around 0.01 and 0.02 for all four configurations. This indicates that AlphaVB is capable of accurately modeling sparsity, and at the same time its decent TPR performance further showcases its balance between finding signal and ignoring noise in the high-dimensional setting. Inspecting why AlphaVB's $\ell_2$ Error rate and MSPE falls slightly short of the best performer within each setting, it remains competitive regardless without any anomaly. On the other hand, AlphaSVB does not work well for any of the settings, forming a significant gap between it and other methods. We posit that the sparse high-dimensional setting coupled with our model specification is not a particularly good fit for direct Monte Carlo optimization, as we derived in \ref{S:mc_vb}. Nonetheless, no single method dominates across all simulations or settings. Therefore, AlphaVB along with other three state-of-the-art methods form a complementary set of tools for practitioners who may flexibly choose the appropriate method based on the needed configuration.

While there is no clear trend when we compare the four simulation settings, configuration (iii) with its high sparsity presents the best performance for many methods across settings. Notably, AlphaVB performs particularly well under this setting, which is not surprising given its good FDR. At the same time, AlphaVB achieves a TPR of 0.91, which is among the best. The other three configurations, on the other hand, do not have a clear difference from the baseline. Our recommendation thus stays the same, and when datasets are particularly sparse, AlphaVB is a compelling method for its good performance across the board.

\begin{table}[ht]
    \centering
    \scriptsize 
    \renewcommand{\arraystretch}{1.1}
    \begin{tabular}{llcccc}
    \toprule
    \textbf{Metric} & \textbf{Method} & \textbf{(i)} & \textbf{(ii)} & \textbf{(iii)} & \textbf{(iv)} \\
    \midrule
    
    \multirow{3}{*}{$\ell_2$ Error}
      & AlphaVB & 0.73 $\pm$ 1.07 & 0.40 $\pm$ 0.06 & 0.19 $\pm$ 0.08 & \textbf{0.30 $\pm$ 0.06} \\
      & AlphaSVB & 3.32 $\pm$ 1.02 & 8.07 $\pm$ 1.44 & 2.10 $\pm$ 1.25 & 4.77 $\pm$ 1.34 \\
      & sparsevb   & 0.51 $\pm$ 0.22 & 0.42 $\pm$ 0.07 & 0.21 $\pm$ 0.12 & 0.34 $\pm$ 0.07 \\
      & spikeslab  & 1.04 $\pm$ 0.56 & 1.95 $\pm$ 0.53 & 0.39 $\pm$ 0.15 & 0.71 $\pm$ 0.24 \\
      & varbvs   & \textbf{0.41 $\pm$ 0.13} & \textbf{0.39 $\pm$ 0.06} & \textbf{0.18 $\pm$ 0.08} & 0.31 $\pm$ 0.06 \\
      & lasso   & 0.82 $\pm$ 0.13  & 0.87 $\pm$ 0.08 & 0.46 $\pm$ 0.08 & 0.62 $\pm$ 0.07 \\
    \midrule
    
    \multirow{3}{*}{FDR} 
      & AlphaVB & 0.02 $\pm$ 0.05 & 0.01 $\pm$ 0.02 & 0.02 $\pm$ 0.06 & \textbf{0.01 $\pm$ 0.02} \\
      & AlphaSVB & 0.10 $\pm$ 0.12 & 0.30 $\pm$ 0.09 & 0.14 $\pm$ 0.20 & 0.03 $\pm$ 0.05\\
      & sparsevb   & 0.09 $\pm$ 0.15 & 0.03 $\pm$ 0.04 & 0.05 $\pm$ 0.14 & 0.04 $\pm$ 0.06 \\
      & spikeslab  & 0.67 $\pm$ 0.11 & 0.74 $\pm$ 0.05 & 0.73 $\pm$ 0.15 & 0.67 $\pm$ 0.07 \\
      & varbvs   & \textbf{0.01 $\pm$ 0.03} & \textbf{0.01 $\pm$ 0.01} & \textbf{0.01 $\pm$ 0.04} & \textbf{0.01 $\pm$ 0.02} \\
      & lasso   & 0.74 $\pm$ 0.07 & 0.77 $\pm$ 0.04 & 0.82 $\pm$ 0.12  & 0.74 $\pm$ 0.06\\
    \midrule
    
    \multirow{3}{*}{TPR}
      & AlphaVB & 0.81 $\pm$ 0.29 & 0.94 $\pm$ 0.05 & 0.91 $\pm$ 0.14 & 0.93 $\pm$ 0.06 \\
      & AlphaSVB & 0.52 $\pm$ 0.15 & 0.50 $\pm$ 0.07 & 0.52 $\pm$ 0.19 & 0.53 $\pm$ 0.10 \\
      & sparsevb   & 0.90 $\pm$ 0.10 & 0.94 $\pm$ 0.04 & 0.92 $\pm$ 0.13 & 0.94 $\pm$ 0.05 \\
      & spikeslab  & 0.82 $\pm$ 0.12 & 0.83 $\pm$ 0.05 & 0.86 $\pm$ 0.15 & 0.91 $\pm$ 0.06 \\
      & varbvs   & 0.89 $\pm$ 0.10 & 0.93 $\pm$ 0.04 & 0.92 $\pm$ 0.13 & 0.93 $\pm$ 0.05 \\
      & lasso   & \textbf{0.94 $\pm$ 0.08}  & \textbf{0.96 $\pm$ 0.03} & \textbf{0.97 $\pm$ 0.07} & \textbf{0.97 $\pm$ 0.04}\\
    \midrule
    
    \multirow{3}{*}{MSPE}
      & AlphaVB & 1.17 $\pm$ 0.9 & 0.92 $\pm$ 0.03 & 0.97 $\pm$ 0.04 & 0.94 $\pm$ 0.04 \\
      & AlphaSVB & 3.15 $\pm$ 0.89 & 7.36 $\pm$ 1.41 & 2.21 $\pm$ 1.10 & 4.70 $\pm$ 1.34 \\
      & sparsevb   & 0.83 $\pm$ 0.12 & 0.87 $\pm$ 0.05 & 0.95 $\pm$ 0.07 & 0.91 $\pm$ 0.05 \\
      & spikeslab  & 0.99 $\pm$ 0.19 & 1.37 $\pm$ 0.25 & 0.94 $\pm$ 0.08 & 0.97 $\pm$ 0.11 \\
      & varbvs   & 0.96 $\pm$ 0.09 & 0.96 $\pm$ 0.04 & 0.99 $\pm$ 0.05 & 0.97 $\pm$ 0.05 \\
      & lasso  & \textbf{0.78 $\pm$ 0.16} & \textbf{0.75 $\pm$ 0.11} & \textbf{0.89 $\pm$ 0.14} & \textbf{0.87 $\pm$ 0.08} \\
    \bottomrule
    \end{tabular} \label{tab:performance_benchmark}
    \caption{A comparison of Sparse High-Dimensional Regression methods using four simulation configurations. All metrics are averages across 100 simulation repeats with the standard deviation as error bounds. The best performance within each setting is presented in bold.}
\end{table}

\subsection{Selection of $\alpha$ Values} \label{S:results_alpha}

One of the unique advantages of AlphaVB and AlphaSVB is that users have the ability to choose a desired $\alpha$ value for a divergence function other than the asymptotic KL divergence. The choice, we argue, is critical in both the desired effect and the downstream application. Therefore, we evaluated the performance of our methods across different $\alpha$ values. First, we considered the performance of AlphaVB as we vary $\alpha$. As mentioned in Section \ref{S:cavi}, we consider only $\alpha>1$ for AlphaVB, and we thus tested $\alpha \in [1.01, 1.1, 1.2, 1.3, 1.5, 2, 3, 5, 100]$. Small $\alpha$ values are closer to KL divergence, whereas $\alpha=100$ represents a relatively extreme configuration. We found that small $\alpha$ values, especially $\alpha=1.01$, work the best for AlphaVB across all metrics, except FDR (Table~\ref{tab:alphasvb_alpha_large}). Interestingly, FDR favors particularly large $\alpha$ values, suggesting that $\alpha$ directly affects the estimation of sparsity. However, using a large $\alpha$ value's tradeoff is severe: a perfect FDR with $\alpha=5$ and $\alpha=100$ dramatically decreases the AlphaVB's performance across all other metrics. Conversely, AlphaVB still achieves excellent, though not perfect, FDR when we select $\alpha=1.01$. To balance the overall performance with FDR, we thus set $\alpha=1.01$ as the default.

AlphaSVB, on the other hand, does not have same restriction on $\alpha$ values as AlphaVB. Therefore, we tested $\alpha \in [0.01, 0.1, 0.25, 0.5, 0.9]$ (Table~\ref{tab:alphasvb_alpha_small}) along with the standard set of $\alpha$ values (Table~\ref{tab:alphasvb_alpha_large}). Through these two benchmarks, a few trends are obvious. First, AlphaSVB has significantly worse performance overall than AlphaVB as previously discussed. Second, there is no clear trend regarding an optimal $\alpha$ value, as seen for AlphaVB. In fact, $\alpha>1$ is suboptimal in general, and as shown in \Cref{tab:alphasvb_alpha_large}, the optimal setting for each configuration varies considerably. Interestingly, AlphaSVB's performance overall is much better when $\alpha<1$ (\Cref{tab:alphasvb_alpha_large}). While quite a few reasonable $\alpha$ values achieve good performance, especially for metrics such as TPR, $\alpha=0.9$ is the most consistent overall, hence motivating our choice as the default. As a general guiding principle for practical implementations of AlphaVB and AlphaSVB, our observation is that $\alpha$ values close to 1 is often ideal, but again, our exact recommendation depends on the configuration.

\begin{table}[ht]
\centering
\resizebox{\textwidth}{!}{%
\begin{tabular}{llcccc | llcccc}
\toprule
\multicolumn{6}{c|}{$\ell_2$ Error} & \multicolumn{6}{c}{FDR} \\
\cmidrule(lr){1-6}\cmidrule(lr){7-12}
 & \textbf{($\alpha$)} & \textbf{(i)} & \textbf{(ii)} & \textbf{(iii)} & \textbf{(iv)} 
& & \textbf{($\alpha$)} & \textbf{(i)} & \textbf{(ii)} & \textbf{(iii)} & \textbf{(iv)} \\
\midrule
& $\alpha=1.01$ & \textbf{0.73 $\pm$ 1.07} &\textbf{ 0.40 $\pm$ 0.06} & \textbf{0.19 $\pm$ 0.08} & \textbf{0.30 $\pm$ 0.06}
& & $\alpha=1.01$ & 0.02 $\pm$ 0.05 & 0.01 $\pm$ 0.02 & 0.02 $\pm$ 0.06 & 0.01 $\pm$ 0.02 \\
& $\alpha=1.10$ & 0.74 $\pm$ 1.07 & 0.42 $\pm$ 0.06 & \textbf{0.19 $\pm$ 0.08} & \textbf{0.30 $\pm$ 0.06} 
& & $\alpha=1.10$ & 0.02 $\pm$ 0.05 & 0.01 $\pm$ 0.02 & 0.01 $\pm$ 0.05 & 0.01 $\pm$ 0.03 \\
& $\alpha=1.20$ & 0.84 $\pm$ 1.11 & 10.84 $\pm$ 1.33 & 0.22 $\pm$ 0.30 & 2.77 $\pm$ 2.47 
& & $\alpha=1.20$ & 0.04 $\pm$ 0.13 & 0.95 $\pm$ 0.02 & 0.02 $\pm$ 0.11 & 0.41 $\pm$ 0.35 \\
& $\alpha=1.30$ & 2.68 $\pm$ 1.85 & 10.67 $\pm$ 0.97 & 0.33 $\pm$ 0.63 & 6.34 $\pm$ 1.66 
& & $\alpha=1.30$ & 0.39 $\pm$ 0.34 & 0.92 $\pm$ 0.10 & 0.05 $\pm$ 0.17 & 0.88 $\pm$ 0.12 \\
& $\alpha=1.50$ & 4.54 $\pm$ 1.40 & 11.02 $\pm$ 0.81 & 1.37 $\pm$ 1.44 & 7.04 $\pm$ 1.27 
& & $\alpha=1.50$ & 0.65 $\pm$ 0.30 & 0.15 $\pm$ 0.32 & 0.35 $\pm$ 0.40 & 0.74 $\pm$ 0.21 \\
& $\alpha=2.00$ & 5.11 $\pm$ 1.12 & 11.05 $\pm$ 0.75 & 2.81 $\pm$ 1.33 & 7.69 $\pm$ 0.89 
& & $\alpha=2.00$ & 0.64 $\pm$ 0.28 & \textbf{0.00 $\pm$ 0.00} & 0.63 $\pm$ 0.33 & 0.02 $\pm$ 0.11 \\
& $\alpha=3.00$ & 4.36 $\pm$ 1.19 & 9.83 $\pm$ 1.28 & 3.63 $\pm$ 0.79 & 6.88 $\pm$ 1.06 
& & $\alpha=3.00$ & 0.02 $\pm$ 0.14 & 0.21 $\pm$ 0.41 & \textbf{0.00 $\pm$ 0.00} & 0.05 $\pm$ 0.22 \\
& $\alpha=5.00$ & 5.01 $\pm$ 0.98 & 9.80 $\pm$ 1.31 & 3.72 $\pm$ 0.77 & 6.86 $\pm$ 1.14 
& & $\alpha=5.00$ & \textbf{0.00 $\pm$ 0.00} & \textbf{0.00 $\pm$ 0.00} & \textbf{0.00 $\pm$ 0.00} & \textbf{0.00 $\pm$ 0.00} \\
& $\alpha=100.00$ & 4.77 $\pm$ 1.08 & 9.79 $\pm$ 1.28 & 3.72 $\pm$ 0.77 & 6.87 $\pm$ 1.13 
& & $\alpha=100.00$ & \textbf{0.00 $\pm$ 0.00} & 0.00 $\pm$ 0.01 & \textbf{0.00 $\pm$ 0.00} & \textbf{0.00 $\pm$ 0.00} \\
\bottomrule
\multicolumn{12}{c}{} \\

\toprule
\multicolumn{6}{c|}{TPR} & \multicolumn{6}{c}{MSPE} \\
\cmidrule(lr){1-6}\cmidrule(lr){7-12}
 & \textbf{($\alpha$)} & \textbf{(i)} & \textbf{(ii)} & \textbf{(iii)} & \textbf{(iv)} 
& & \textbf{($\alpha$)} & \textbf{(i)} & \textbf{(ii)} & \textbf{(iii)} & \textbf{(iv)} \\
\midrule
& $\alpha=1.01$ & \textbf{0.81 $\pm$ 0.29} & \textbf{0.94 $\pm$ 0.05} & \textbf{0.91 $\pm$ 0.14} & \textbf{0.93 $\pm$ 0.06}
& & $\alpha=1.01$ & \textbf{1.17 $\pm$ 0.90} & \textbf{0.92 $\pm$ 0.03} & \textbf{0.97 $\pm$ 0.04} & \textbf{0.94 $\pm$ 0.04} \\
& $\alpha=1.10$ & \textbf{0.81 $\pm$ 0.29} & \textbf{0.94 $\pm$ 0.05} & \textbf{0.91 $\pm$ 0.14} & \textbf{0.93 $\pm$ 0.06}
& & $\alpha=1.10$ & 1.18 $\pm$ 0.90 & 0.93 $\pm$ 0.03 & \textbf{0.97 $\pm$ 0.04} & \textbf{0.94 $\pm$ 0.04} \\
& $\alpha=1.20$ & \textbf{0.81 $\pm$ 0.29} & 0.30 $\pm$ 0.08 & \textbf{0.91 $\pm$ 0.14} & 0.79 $\pm$ 0.14
& & $\alpha=1.20$ & 1.22 $\pm$ 0.90 & 5.45 $\pm$ 1.42 & 0.98 $\pm$ 0.09 & 1.91 $\pm$ 0.84 \\
& $\alpha=1.30$ & 0.66 $\pm$ 0.27 & 0.17 $\pm$ 0.05 & 0.89 $\pm$ 0.15 & 0.39 $\pm$ 0.13
& & $\alpha=1.30$ & 1.78 $\pm$ 0.88 & 8.31 $\pm$ 1.04 & 1.00 $\pm$ 0.11 & 4.04 $\pm$ 1.24 \\
& $\alpha=1.50$ & 0.32 $\pm$ 0.19 & 0.02 $\pm$ 0.04 & 0.72 $\pm$ 0.22 & 0.21 $\pm$ 0.09
& & $\alpha=1.50$ & 3.23 $\pm$ 0.98 & 10.80 $\pm$ 1.19 & 1.31 $\pm$ 0.48 & 6.12 $\pm$ 1.30 \\
& $\alpha=2.00$ & 0.16 $\pm$ 0.11 & 0.00 $\pm$ 0.00 & 0.37 $\pm$ 0.21 & 0.01 $\pm$ 0.04
& & $\alpha=2.00$ & 4.43 $\pm$ 0.95 & 11.05 $\pm$ 0.82 & 2.55 $\pm$ 1.01 & 7.79 $\pm$ 0.99 \\
& $\alpha=3.00$ & 0.18 $\pm$ 0.16 & 0.11 $\pm$ 0.09 & 0.03 $\pm$ 0.08 & 0.11 $\pm$ 0.09
& & $\alpha=3.00$ & 4.39 $\pm$ 1.19 & 9.60 $\pm$ 1.34 & 3.77 $\pm$ 0.78 & 6.86 $\pm$ 1.11 \\
& $\alpha=5.00$ & 0.08 $\pm$ 0.15 & 0.10 $\pm$ 0.09 & 0.00 $\pm$ 0.00 & 0.10 $\pm$ 0.10
& & $\alpha=5.00$ & 5.09 $\pm$ 1.03 & 9.58 $\pm$ 1.39 & 3.86 $\pm$ 0.76 & 6.85 $\pm$ 1.18 \\
& $\alpha=100.00$ & 0.10 $\pm$ 0.12 & 0.10 $\pm$ 0.09 & 0.00 $\pm$ 0.00 & 0.10 $\pm$ 0.09
& & $\alpha=100.00$ & 4.82 $\pm$ 1.12 & 9.57 $\pm$ 1.37 & 3.86 $\pm$ 0.76 & 6.86 $\pm$ 1.19 \\
\bottomrule
\end{tabular} \label{tab:alphavb_alpha_large}
}
\caption{Performance of AlphaVB across $\alpha \in [1.01, 1.1, 1.2, 1.3, 1.5, 2, 3, 5, 100]$ and four configurations. All metrics are averages across 100 simulation repeats with the standard deviation as error bounds. The best performance within each setting is presented in bold.}
\end{table}

\begin{table}[ht]
\centering
\resizebox{\textwidth}{!}{%
\begin{tabular}{llcccc | llcccc}
\toprule
\multicolumn{6}{c|}{$\ell_2$ Error} & \multicolumn{6}{c}{FDR} \\
\cmidrule(lr){1-6}\cmidrule(lr){7-12}
 & \textbf{($\alpha$)} & \textbf{(i)} & \textbf{(ii)} & \textbf{(iii)} & \textbf{(iv)} 
& & \textbf{($\alpha$)} & \textbf{(i)} & \textbf{(ii)} & \textbf{(iii)} & \textbf{(iv)} \\
\midrule
& $\alpha=1.01$ & 7.89 $\pm$ 2.87 & 17.25 $\pm$ 3.92 & 10.12 $\pm$ 3.51 & 9.59 $\pm$ 3.11
& & $\alpha=1.01$ & 0.80 $\pm$ 0.06 & 0.84 $\pm$ 0.03 & 0.95 $\pm$ 0.02 & \textbf{0.74 $\pm$ 0.05} \\
& $\alpha=1.10$ & 8.35 $\pm$ 3.49 & 17.06 $\pm$ 3.29 & 10.54 $\pm$ 4.66 & 9.59 $\pm$ 2.71
& & $\alpha=1.10$ & 0.80 $\pm$ 0.05 & 0.85 $\pm$ 0.02 & 0.95 $\pm$ 0.02 & 0.75 $\pm$ 0.06 \\
& $\alpha=1.20$ & 7.90 $\pm$ 2.24 & 17.60 $\pm$ 3.55 & 10.34 $\pm$ 9.26 & 10.25 $\pm$ 3.97
& & $\alpha=1.20$ & 0.79 $\pm$ 0.06 & 0.85 $\pm$ 0.02 & 0.95 $\pm$ 0.02 & 0.74 $\pm$ 0.06 \\
& $\alpha=1.30$ & 8.19 $\pm$ 3.66 & 17.10 $\pm$ 3.12 & 9.96 $\pm$ 2.85 & \textbf{9.40 $\pm$ 2.79}
& & $\alpha=1.30$ & 0.79 $\pm$ 0.06 & \textbf{0.84 $\pm$ 0.02} & 0.95 $\pm$ 0.02 & 0.74 $\pm$ 0.07 \\
& $\alpha=1.50$ & 8.51 $\pm$ 3.04 & 17.30 $\pm$ 3.35 & 9.97 $\pm$ 3.32 & 9.72 $\pm$ 2.81
& & $\alpha=1.50$ & \textbf{0.79 $\pm$ 0.05} & \textbf{0.84 $\pm$ 0.02} & 0.95 $\pm$ 0.02 & 0.76 $\pm$ 0.05 \\
& $\alpha=2.00$ & 8.36 $\pm$ 3.11 & \textbf{16.29 $\pm$ 2.51} & 10.46 $\pm$ 4.90 & 9.78 $\pm$ 3.25
& & $\alpha=2.00$ & 0.80 $\pm$ 0.06 & 0.84 $\pm$ 0.03 & \textbf{0.95 $\pm$ 0.01} & 0.75 $\pm$ 0.06 \\
& $\alpha=3.00$ & \textbf{7.88 $\pm$ 2.71} & 17.07 $\pm$ 3.32 & \textbf{9.53 $\pm$ 2.94} & 9.44 $\pm$ 3.11
& & $\alpha=3.00$ & 0.80 $\pm$ 0.07 & \textbf{0.84 $\pm$ 0.02} & 0.95 $\pm$ 0.02 & \textbf{0.74 $\pm$ 0.05} \\
& $\alpha=5.00$ & 8.90 $\pm$ 3.59 & 16.87 $\pm$ 3.80 & 9.76 $\pm$ 3.25 & 9.72 $\pm$ 2.89
& & $\alpha=5.00$ & 0.80 $\pm$ 0.06 & 0.84 $\pm$ 0.03 & 0.95 $\pm$ 0.02 & \textbf{0.74 $\pm$ 0.05} \\
& $\alpha=100.00$ & 8.21 $\pm$ 2.75 & 17.48 $\pm$ 3.34 & 10.49 $\pm$ 3.85 & 10.12 $\pm$ 4.05
& & $\alpha=100.00$ & 0.80 $\pm$ 0.05 & \textbf{0.84 $\pm$ 0.02} & 0.95 $\pm$ 0.02 & 0.74 $\pm$ 0.06 \\
\bottomrule
\multicolumn{12}{c}{} \\

\toprule
\multicolumn{6}{c|}{TPR} & \multicolumn{6}{c}{MSPE} \\
\cmidrule(lr){1-6}\cmidrule(lr){7-12}
 & \textbf{($\alpha$)} & \textbf{(i)} & \textbf{(ii)} & \textbf{(iii)} & \textbf{(iv)} 
& & \textbf{($\alpha$)} & \textbf{(i)} & \textbf{(ii)} & \textbf{(iii)} & \textbf{(iv)} \\
\midrule
& $\alpha=1.01$ & 0.53 $\pm$ 0.16 & 0.44 $\pm$ 0.08 & 0.62 $\pm$ 0.21 & 0.49 $\pm$ 0.11
& & $\alpha=1.01$ & \textbf{7.32 $\pm$ 2.73} & 16.27 $\pm$ 3.93 & 9.89 $\pm$ 3.45 & 9.27 $\pm$ 3.02 \\
& $\alpha=1.10$ & 0.53 $\pm$ 0.14 & \textbf{0.42 $\pm$ 0.08} & 0.63 $\pm$ 0.20 & 0.48 $\pm$ 0.12
& & $\alpha=1.10$ & 7.87 $\pm$ 3.73 & 16.13 $\pm$ 3.51 & 10.30 $\pm$ 4.65 & 9.29 $\pm$ 2.72 \\
& $\alpha=1.20$ & 0.54 $\pm$ 0.15 & 0.43 $\pm$ 0.08 & \textbf{0.59 $\pm$ 0.20} & 0.48 $\pm$ 0.12
& & $\alpha=1.20$ & 7.49 $\pm$ 2.27 & 16.58 $\pm$ 3.59 & 10.07 $\pm$ 8.76 & 9.95 $\pm$ 4.01 \\
& $\alpha=1.30$ & 0.53 $\pm$ 0.14 & 0.44 $\pm$ 0.07 & 0.60 $\pm$ 0.21 & 0.49 $\pm$ 0.12
& & $\alpha=1.30$ & 7.68 $\pm$ 3.86 & 16.03 $\pm$ 3.20 & 9.72 $\pm$ 2.77 & 9.08 $\pm$ 2.82 \\
& $\alpha=1.50$ & 0.55 $\pm$ 0.16 & 0.43 $\pm$ 0.09 & 0.60 $\pm$ 0.19 & \textbf{0.47 $\pm$ 0.11}
& & $\alpha=1.50$ & 8.04 $\pm$ 2.96 & 16.50 $\pm$ 3.48 & 9.80 $\pm$ 3.43 & 9.35 $\pm$ 2.81 \\
& $\alpha=2.00$ & \textbf{0.52 $\pm$ 0.16} & 0.44 $\pm$ 0.08 & 0.60 $\pm$ 0.21 & 0.48 $\pm$ 0.13
& & $\alpha=2.00$ & 7.90 $\pm$ 3.04 & \textbf{15.48 $\pm$ 2.55} & 10.19 $\pm$ 4.87 & 9.36 $\pm$ 3.26 \\
& $\alpha=3.00$ & \textbf{0.52 $\pm$ 0.16} & 0.44 $\pm$ 0.08 & 0.61 $\pm$ 0.19 & 0.48 $\pm$ 0.11
& & $\alpha=3.00$ & 7.46 $\pm$ 2.68 & 16.20 $\pm$ 3.34 & \textbf{9.35 $\pm$ 2.89} & \textbf{9.07 $\pm$ 3.12} \\
& $\alpha=5.00$ & \textbf{0.52 $\pm$ 0.16} & 0.44 $\pm$ 0.08 & 0.60 $\pm$ 0.20 & 0.48 $\pm$ 0.12
& & $\alpha=5.00$ & 8.04 $\pm$ 3.58 & 16.00 $\pm$ 3.82 & 9.61 $\pm$ 3.23 & 9.35 $\pm$ 2.91 \\
& $\alpha=100.00$ & 0.53 $\pm$ 0.15 & 0.44 $\pm$ 0.07 & 0.59 $\pm$ 0.21 & 0.48 $\pm$ 0.12
& & $\alpha=100.00$ & 7.51 $\pm$ 2.76 & 16.49 $\pm$ 3.36 & 10.28 $\pm$ 3.80 & 9.78 $\pm$ 4.04 \\
\bottomrule
\end{tabular} \label{tab:alphasvb_alpha_large}
}
\caption{Performance of AlphaSVB across nine $\alpha$ values, $\alpha \in \{1.01, 1.10, 1.20, 1.30, 1.50, 2.00, 3.00, 5.00, 100.00\}$. All metrics are averages across 100 simulation repeats with the standard deviation as error bounds. The best performance within each setting is presented in bold.}
\end{table}

\begin{table}[ht]
\centering
\resizebox{\textwidth}{!}{%
\begin{tabular}{llcccc | llcccc}
\toprule
\multicolumn{6}{c|}{$\ell_2$ Error} & \multicolumn{6}{c}{FDR} \\
\cmidrule(lr){1-6}\cmidrule(lr){7-12}
 & \textbf{($\alpha$)} & \textbf{(i)} & \textbf{(ii)} & \textbf{(iii)} & \textbf{(iv)}
 & & \textbf{($\alpha$)} & \textbf{(i)} & \textbf{(ii)} & \textbf{(iii)} & \textbf{(iv)} \\
\midrule
& $\alpha=0.01$ & 3.54 $\pm$ 1.14 & 8.07 $\pm$ 1.64 & 2.36 $\pm$ 0.89 & 5.13 $\pm$ 1.21
& & $\alpha=0.01$ & 0.10 $\pm$ 0.13 & 0.31 $\pm$ 0.09 & 0.15 $\pm$ 0.19 & 0.03 $\pm$ 0.06 \\
& $\alpha=0.10$ & 3.69 $\pm$ 0.98 & 8.07 $\pm$ 1.54 & 2.42 $\pm$ 0.97 & 5.07 $\pm$ 1.37
& & $\alpha=0.10$ & 0.11 $\pm$ 0.16 & 0.31 $\pm$ 0.10 & 0.16 $\pm$ 0.21 & 0.04 $\pm$ 0.06 \\
& $\alpha=0.25$ & 3.71 $\pm$ 1.08 & \textbf{8.01 $\pm$ 1.61} & 2.33 $\pm$ 0.98 & 5.06 $\pm$ 1.26
& & $\alpha=0.25$ & 0.12 $\pm$ 0.14 & 0.32 $\pm$ 0.10 & \textbf{0.10 $\pm$ 0.17} & 0.03 $\pm$ 0.06 \\
& $\alpha=0.50$ & 3.50 $\pm$ 1.05 & 8.04 $\pm$ 1.45 & 2.38 $\pm$ 1.11 & 5.16 $\pm$ 1.24
& & $\alpha=0.50$ & 0.13 $\pm$ 0.14 & 0.33 $\pm$ 0.09 & 0.15 $\pm$ 0.22 & 0.04 $\pm$ 0.06 \\
& $\alpha=0.90$ & \textbf{3.32 $\pm$ 1.02} & 8.07 $\pm$ 1.44 & \textbf{2.10 $\pm$ 1.25} & \textbf{4.77 $\pm$ 1.34}
& & $\alpha=0.90$ & \textbf{0.10 $\pm$ 0.12} & \textbf{0.30 $\pm$ 0.09} & 0.14 $\pm$ 0.20 & \textbf{0.03 $\pm$ 0.05} \\
\bottomrule
\multicolumn{12}{c}{} \\

\toprule
\multicolumn{6}{c|}{TPR} & \multicolumn{6}{c}{MSPE} \\
\cmidrule(lr){1-6}\cmidrule(lr){7-12}
 & \textbf{($\alpha$)} & \textbf{(i)} & \textbf{(ii)} & \textbf{(iii)} & \textbf{(iv)}
 & & \textbf{($\alpha$)} & \textbf{(i)} & \textbf{(ii)} & \textbf{(iii)} & \textbf{(iv)} \\
\midrule
& $\alpha=0.01$ & 0.46 $\pm$ 0.16 & 0.52 $\pm$ 0.08 & \textbf{0.50 $\pm$ 0.20} & 0.50 $\pm$ 0.13
& & $\alpha=0.01$ & 3.34 $\pm$ 1.03 & 7.35 $\pm$ 1.61 & 2.40 $\pm$ 0.76 & 4.97 $\pm$ 1.15 \\
& $\alpha=0.10$ & 0.47 $\pm$ 0.17 & 0.50 $\pm$ 0.08 & 0.50 $\pm$ 0.21 & 0.49 $\pm$ 0.12
& & $\alpha=0.10$ & 3.49 $\pm$ 0.92 & 7.32 $\pm$ 1.49 & 2.49 $\pm$ 0.92 & 4.93 $\pm$ 1.33 \\
& $\alpha=0.25$ & \textbf{0.46 $\pm$ 0.15} & 0.49 $\pm$ 0.08 & 0.52 $\pm$ 0.23 & 0.50 $\pm$ 0.12
& & $\alpha=0.25$ & 3.50 $\pm$ 0.97 & \textbf{7.23 $\pm$ 1.58} & 2.40 $\pm$ 0.88 & 4.92 $\pm$ 1.20 \\
& $\alpha=0.50$ & 0.48 $\pm$ 0.14 & \textbf{0.49 $\pm$ 0.07} & 0.50 $\pm$ 0.22 & \textbf{0.49 $\pm$ 0.11}
& & $\alpha=0.50$ & 3.34 $\pm$ 0.93 & 7.25 $\pm$ 1.40 & 2.44 $\pm$ 1.00 & 4.99 $\pm$ 1.20 \\
& $\alpha=0.90$ & 0.52 $\pm$ 0.15 & 0.50 $\pm$ 0.07 & 0.52 $\pm$ 0.19 & 0.53 $\pm$ 0.10
& & $\alpha=0.90$ & \textbf{3.15 $\pm$ 0.89} & 7.36 $\pm$ 1.41 & \textbf{2.21 $\pm$ 1.10} & \textbf{4.70 $\pm$ 1.34} \\
\bottomrule
\end{tabular}%
} \label{tab:alphasvb_alpha_small}
\caption{Performance of AlphaSVB across $\alpha \in \{0.01, 0.10, 0.25, 0.50, 0.90\}$. All metrics are averages across 100 simulation repeats with the standard deviation as error bounds. The best performance within each setting is presented in bold.}
\end{table}

\subsection{Efficient Computation and Implementation}

In this paper, we focused on VI due to its computational efficiency compared to other Bayesian methods such as MCMC. Our simulations confirm this advantage. The runtime of AlphaVB is on the order of seconds to minutes, even with a pure R implementation and for high-dimensional datasets. On the other hand, Ray and Szabó's C++ implementation \citep{ray2022variational} is naturally faster, but this primarily reflects language level optimization rather than methodological inefficiency. We recognize that further runtime improvements for AlphaVB can be achieved through a hybrid implementation to enhance code efficiency. 

\section{Discussion and Future Work}
\label{S:discussion}

In this work, we proposed AlphaVB and AlphaSVB, two variational-inference-based methods that tackle the sparse high-dimensional regression problem. Our results showed promising performance that is competitive against many state-of-the art methods. Specifically, AlphaVB based on a CAVI algorithm with Delta method approximation is a compelling and competitive choice for future applications. On the other hand, AlphaSVB serves as a case in point for the implementation of MC approximation, which is conceptually simpler to implement; however, AlphaSVB's performance deficiencies also highlight the challenges with convergence and accuracy in this setting. By rigorously considering Rényi's $\alpha$-divergence in the context of high-dimensional linear regression, our study serves as a unique pilot for the wider adoption of VI with $\alpha$-divergence to more settings.

The choice of $\alpha$ in Rényi's $\alpha$-divergence is critical in ensuring both meaningful interpretation of the divergence function and good performance in VI approximation. In this work, we have demonstrated that both AlphaVB and AlphaSVB demonstrate various degrees of performance gain by carefully selecting $\alpha$ values. We found that AlphaVB is noticeably more robust against $\alpha$ selection, with small $\alpha\approx 1+\epsilon$ yielding reasonable results. Our results highlight the consistent performance with $\alpha=1.01$. AlphaSVB, on the other hand, requires more attention in the tuning of its hyperparameter. Our benchmark results provide a good starting point for practitioners who wish to adopt our method for practical applications. Yet, we recognize that there is no one-size-fits-all solution: there exists a high degree of complementarity between the metrics used and the simulation configurations. Therefore, we encourage the finetuning of $\alpha$ based on the specific use case, and for similar endeavors involving adapting $\alpha$-divergence to broader VI models, a similar benchmark is thus necessary.

Going beyond AlphaVB and AlphaSVB, we recognize the vast potential for using this work as a foundation for future endeavors. First, there is great interest in generalized linear models (GLMs) for the sparse high-dimensional setting. A natural direction is to extend our work by considering GLM applications, such as logistic regression. The change of a link function is oftentimes nontrivial, and this extension will allow us to more flexibly model different types of outcomes with applications in areas such as bioinformatics. Secondly, the performance of AlphaSVB as derived in this paper warrants a study of MC approximation of $\alpha$ divergence to different settings beyond those presented in \cite{li2016renyi}. A closer look at the empirical as well as theoretical properties of such MC approximation is of interest, as its simplicity in implementation is appealing. Third, the computational efficiency of AlphaVB and AlphaSVB can be further improved. While our CAVI is sufficiently fast for many use cases, the adaptation of large-scale datasets with large $n$ and even larger $p$ renders efficiency critical. For example, our algorithm can be easily adapted to use hybrid implementation via a C++ core and R wrapper using \texttt{Rcpp} \citep{eddelbuettel2011rcpp}. Further, given that the CAVI relies on numerical optimization, models with closed-form solutions can also be the focus in the future. While AlphaVB and AlphaSVB have paved the way, our hope is that this paper will generate more interest in the combination of $\alpha$-divergence, sparse high-dimensional linear regression, and VI.

\newpage
\bibliographystyle{apalike}
\bibliography{vi}  

\begin{appendix}
\section{CAVI Update Derivations}
\subsection{Deriving $\mu$ and $\sigma$ using the multivariate Delta theorem}
\label{A:delta}

We begin with the fact that
\[
\frac{1}{2}\|Y - X\theta\|_2^2
= \frac{1}{2} Y^\top Y - Y^\top X\theta 
+ \frac{1}{2}\theta^\top X^\top X\theta.
\]

\noindent Consider the expression
\[
\exp\!\left\{(\alpha - 1)\!\left(
\frac{1}{2}(Y - X\theta)^\top (Y - X\theta)
+ \lambda |\theta_i|
- \frac{(\theta_i - \mu_i)^2}{2\sigma_i^2}
- \log \sigma_i
\right)\!\right\},
\]
which can be rewritten as
\[
\exp\!\left\{\frac{\alpha-1}{2}Y^\top Y\right\}
\cdot
\exp\!\left\{(\alpha-1)\!\left(
- Y^\top X\theta
+ \frac{1}{2}\theta^\top X^\top X\theta
+ \lambda |\theta_i|
- \frac{(\theta_i - \mu_i)^2}{2\sigma_i^2}
- \log \sigma_i
\right)\right\}.
\]

\noindent
The first factor,
\(
\exp\{\frac{\alpha-1}{2}Y^\top Y\},
\)
is independent of $\theta_i$. We now isolate the $\theta_i$ terms:
\[
- Y^\top X\theta
= - (Y^\top X)_i\theta_i
+ \text{terms independent of }\theta_i,
\]
and
\[
\frac{1}{2}\theta^\top X^\top X\theta
= \theta_i \sum_{j\neq i}\theta_j (X^\top X)_{ji}
+ \frac{1}{2}\theta_i^2 (X^\top X)_{ii}
+ \text{terms independent of }\theta_i.
\]

\noindent Thus,
\begin{align*}
\mathbb{E}_{\mu,\sigma,\gamma \mid z_i = 1}
\!\left[
\left(
\frac{
\frac{d P_{\mu_{-i},\sigma_{-i},\gamma_{-i} \mid z_i=1}
\otimes \nm(\mu_i,\sigma_i^2)}
{d\pi_{-i} \otimes \bar{w}_i \mathrm{Lap}(\lambda)}
}{
D_{\pi}^{-1}
\exp\!\left\{
\frac{1}{2}
\left( -\frac{\|Y - X\theta\|_2^2}{2} \right)
\right\}
}
\right)^{\alpha - 1}
\right]
\end{align*}
equals
\begin{align*}
C \cdot
\mathbb{E}_{\mu,\sigma,\gamma \mid z_i = 1}
\left[
\exp\!\left\{
(\alpha - 1)\!\left(
\frac{1}{2}\|Y - X\theta\|_2^2
+ \lambda |\theta_i|
- \frac{(\theta_i - \mu_i)^2}{2\sigma_i^2}
\right)
\right\}
\left(\frac{1}{\sigma_i}\right)^{\alpha - 1}
\right],
\end{align*}
and maximizing is equivalent to
\[
\scalebox{0.8}{$
\mathbb{E}_{\mu,\sigma,\gamma \mid z_i = 1}\!\left[
\exp\!\left\{
(\alpha - 1)\!
\left(
- (Y^\top X)_i\theta_i
+ \frac{1}{2}\theta_i^2 (X^\top X)_{ii}
+ \theta_i \sum_{j\neq i}\theta_j (X^\top X)_{ji}
+ \lambda |\theta_i|
- \frac{(\theta_i - \mu_i)^2}{2\sigma_i^2}
- \log \sigma_i
\right)
\right\}
\right]
$}
\]

\medskip

\noindent
For CAVI updates for $\gamma_i$, we want to maximize
\begin{align*}
\mathbb{E}_{\mu,\sigma,\gamma}
\Bigg[
\exp\!\Bigg\{
(\alpha - 1)\Bigg(
&- (Y^\top X)_i \theta_i 
+ \frac{1}{2}\theta_i^2 (X^\top X)_{ii}
+ \theta_i \sum_{j\neq i}(X^\top X)_{ji}\theta_j \\
&\quad
+ z_i\!\left(
\log\frac{\sqrt{2}}{\sqrt{\pi}\sigma_i\lambda}
- \frac{(\theta_i - \mu_i)^2}{2\sigma_i^2}
+ \frac{\theta_i^2}{2}\tau
+ \log\frac{\gamma_i}{\bar{w}_i}
\right) \\
&\quad
+ (1 - z_i)\log\frac{1 - \gamma_i}{1 - \bar{w}_i}
\Bigg)
\Bigg\}
\Bigg].
\end{align*}

\begin{theorem}[Multivariate Delta Theorem]
Let $f: \mathbb{R}^k \to \mathbb{R}$ be twice continuously differentiable in a neighborhood of $\mathbb{E}[V]$ where $V$ is a random variable in $\mathbb{R}^k$. Then,
\[
\mathbb{E}[V] \approx f(\mathbb{E}[V]) + \frac{1}{2} \operatorname{tr}(\nabla^2) f(\mathbb{E}[V]) \cdot \operatorname{Cov}(V).
\]
\end{theorem}

\noindent Strictly speaking, the theorem requires twice continuous differentiability, which holds when we apply the approximation
\(
|\theta_i| \approx \sqrt{\theta_i^2 + \varepsilon}
\)
where $\varepsilon \approx 0$. Thus, we are working with the multivariate function $g : \mathbb{R}^p \to \mathbb{R}$ defined by
\begin{align*}
g(\theta)
&= \exp \Biggl\{ (\alpha-1) \Biggl(
- (Y^\top X)_i \theta_i 
+ \frac{1}{2} \theta_i^2 (X^\top X)_{ii}
+ \theta_i \sum_{j \neq i} \theta_j (X^\top X)_{ji} \\
&\quad + \lambda \sqrt{\theta_i^2 + \varepsilon}
- \frac{(\theta_i - \mu_i)^2}{2\sigma_i^2}
- \log \sigma_i
\Biggr) \Biggr\}.
\end{align*}

\medskip

\noindent
Note that $\mathbb{E}[\theta_i \mid z_i = 1] = \mu_i$. Furthermore,
\begin{align*}
\mathbb{E}[\theta_j \mid i \neq j]
&= \mathbb{E}\!\left[ \gamma_j N(\mu_j, \sigma_j^2) + (1 - \gamma_j) \delta_0 \right] \\
&= \gamma_j \, \mathbb{E}[N(\mu_j, \sigma_j^2)] + (1 - \gamma_j)\cdot 0 \\
&= \gamma_j \mu_j.
\end{align*}

\noindent
In general,
\[
\mathbb{E}[\theta]_j =
\begin{cases}
\mu_i, & \text{if } j = i, \\[1ex]
\gamma_j \mu_j, & \text{otherwise}.
\end{cases}
\]

\bigskip

\noindent
Let us compute the covariance. Let $n,m \in [p]$. If $m \neq n$, then
\[
\mathrm{Cov}(\theta_n, \theta_m) = 0,
\]
since $\theta_n$ and $\theta_m$ are independent. Now consider
\[
\mathrm{Var}[\theta_n]
= \mathbb{E}[\theta_n^2] - (\mathbb{E}[\theta_n])^2.
\]

\medskip

\noindent
If $n = i$, then
\[
\mathrm{Var}[\theta_i] = \sigma_i^2
\qquad \text{(since we conditioned on $z_i = 1$)}.
\]
If $n \neq i$, then
\[
\mathrm{Var}[\theta_n]
= \mathbb{E}[\theta_n^2] - (\gamma_n \mu_n)^2.
\]

\medskip

\noindent
It can be shown that
\begin{align*}
\mathbb{E}[\theta_n^2]
&= \int \theta_n^2 \left[ \gamma_n N(\theta_n; \mu_n, \sigma_n^2)
+ (1 - \gamma_n)\delta_0 \right] d\theta_n \\
&= \gamma_n \int \theta_n^2 N(\theta_n; \mu_n, \sigma_n^2) d\theta_n
 + (1 - \gamma_n) \int \theta_n^2 \delta_0 \, d\theta_n \\
&= \gamma_n (\mu_n^2 + \sigma_n^2)
 + (1 - \gamma_n)(0^2 + 0^2) \\
&= \gamma_n (\mu_n^2 + \sigma_n^2).
\end{align*}

\medskip

\noindent
Thus,
\[
\mathrm{Var}[\theta_n \mid n \neq i]
= \gamma_n (\mu_n^2 + \sigma_n^2)
- \gamma_n^2 \mu_n^2
= (\gamma_n - \gamma_n^2)\mu_n^2 + \gamma_n \sigma_n^2.
\]

\bigskip

\noindent
Therefore,
\[
\bigl[\mathrm{Cov}(\theta)\bigr]_{nm} =
\begin{cases}
\sigma_i^2, & n = m = i, \\[2ex]
\gamma_n(1 - \gamma_n)\mu_n^2 + \gamma_n \sigma_n^2, & n = m \neq i, \\[2ex]
0, & n \neq m.
\end{cases}
\]

\subsection{Deriving $\gamma$ updates}
\label{A:jensen}

Since $x \mapsto \exp\{a x\}$ is strictly convex for all $a \in \mathbb{R}$, Jensen's inequality implies that
\begin{equation*}
\mathbb{E}_{\mu, \sigma, \gamma} \left[ 
\exp \left\{ (\alpha - 1) \, \psi(\theta, z_i) \right\} 
\right]
\geq
\exp \left\{ (\alpha - 1) \, \mathbb{E}_{\mu, \sigma, \gamma} \left[ \psi(\theta, z_i) \right] \right\},
\end{equation*}
where
\begin{align*}
\psi(\theta, z_i) &= 
- (Y^\top X)_i \theta_i 
+ \frac{1}{2} \theta_i^2 (X^\top X)_{ii} 
+ \theta_i \sum_{j \neq i} (X^\top X)_{ji} \theta_j \\
&+ \mathbb{I}_{\{z_i = 1\}}\! \biggl(
\log \frac{\sqrt{2}}{\sqrt{\pi}\sigma_i \lambda} 
- \frac{(\theta_i - \mu_i)^2}{2\sigma_i^2} 
+ \lambda |\theta_i| 
+ \log \frac{\gamma_i}{\bar{w}_i}
\biggr) \\
&+ \mathbb{I}_{\{z_i = 0\}}\! \log \frac{1-\gamma_i}{1-\bar{w}_i}.
\end{align*}
 Using the monotonicity of $x \mapsto \exp \{ax\}$ reduces the minimization problem to a maximization of $\mathbb{E}_{\mu, \sigma, \gamma} \left[ \psi(\theta, z_i) \right] $.
Recall that $Z \sim \nm(\mu, \sigma^2)$. Then, $|Z|$, also known as a folded Gaussian variable, has 
\[\mathbb{E}[|Z|] = \sigma \sqrt{\frac{2}{\pi}} \exp \biggl\{ -\frac{\mu^2}{2\sigma^2}\biggr\} + \mu (1- 2\Phi(-\frac{\mu}{\sigma})),\]
where $\Phi$ is the standard normal CDF.
 Note that $\theta_i|z_i = 1 \sim \nm(\mu_i, \sigma_i^2)$. Therefore, $\mathbb{E}_{\mu, \sigma, \gamma} [-\frac{(\theta_i - \mu_i)^2}{2\sigma_i^2}] = -\frac{1}{2}$.
\\
\\
 Evaluating the expectation of the term $\mathbb{I}_{\{z_i = 1\}}\log \frac{\gamma_i}{\bar{w}_i}
+ \mathbb{I}_{\{z_i = 0\}}\! \log \frac{1-\gamma_i}{1-\bar{w}_i}$, we get
\[\gamma_i\log \frac{\gamma_i}{\bar{w}_i}
+ (1-\gamma_i) \log \frac{1-\gamma_i}{1-\bar{w}_i}\]
since $z_i \sim \text{Bernouilli}(\gamma_i)$. 
\\
\\
Taking the expectation of the term $- (Y^\top X)_i \theta_i 
+ \frac{1}{2} \theta_i^2 (X^\top X)_{ii} 
+ \theta_i \sum_{j \neq i} (X^\top X)_{ji} \theta_j$, we get
\[ - (Y^\top X)_i \mu_i 
+ \frac{1}{2} (\mu_i^2 + \sigma_i^2) (X^\top X)_{ii} 
+ \mu_i \sum_{j \neq i} (X^\top X)_{ji} \gamma_j \mu_j.\]
As for the expectation of the term $ \mathbb{I}_{\{z_i = 1\}}\! \biggl(
\log \frac{\sqrt{2}}{\sqrt{\pi}\sigma_i \lambda} 
- \frac{(\theta_i - \mu_i)^2}{2\sigma_i^2} 
+ \lambda |\theta_i| 
\biggr)$, we get
\[\gamma_i \biggl(\log \frac{\sqrt{2}}{\sqrt{\pi}\sigma_i \lambda} - \frac{1}{2} + \lambda \cdot \sigma_i \sqrt{\frac{2}{\pi}} \exp \biggl\{ -\frac{\mu_i^2}{2\sigma_i^2}\biggr\} + \lambda \cdot \mu_i \biggl(1- 2\Phi\biggl(-\frac{\mu_i}{\sigma_i}\biggr)\biggr) \biggr).\]

\end{appendix}

\end{document}